# A comparison of regression models for static and dynamic prediction of a prognostic outcome during admission in electronic health care records

GAO Shan[1], ALBU Elena[1], PUTTER Hein[2], Stijnen Pieter[3], RADEMAKERS Frank[4], COSSEY Veerle[1,5], DEBAVEYE Yves[6], JANSSENS Christel[7], VAN CALSTER Ben[1,2,8,*], WYNANTS Laure[1,8,9]

1 Department of Development and Regeneration, KU Leuven, Belgium

2 Department of Biomedical Data Sciences, Leiden University Medical Center, the Netherlands

3 Management Information Reporting Department, University Hospitals Leuven, Belgium

4 Faculty of Medicine, KU Leuven, Belgium

5 Department of Infection Control and Prevention, University Hospitals Leuven, Belgium

6 Department of Cellular and Molecular Medicine, University Hospitals Leuven, Belgium

7 Nursing PICC team, University Hospitals Leuven, Belgium

8 Unit for Health Technology Assessment Research (LUHTAR), KU Leuven, Belgium

9 School for Public Health and Primary Care, Maastricht University, the Netherlands

Corresponding author. Herestraat 49 - box 805, 3000 Leuven, Belgium. Tel: 003216377788; *E-mail address:* ben.vancalster@kuleuven.be

# Abstract

**Objective** Hospitals register information in the electronic health records (EHR) continuously until discharge or death. As such, there is no censoring for in-hospital outcomes. We aimed to compare different dynamic regression modeling approaches to predict central line-associated bloodstream infections (CLABSI) in EHR while accounting for competing events precluding CLABSI.

**Materials and Methods** We analyzed data from 30,862 catheter episodes at University Hospitals Leuven from 2012 and 2013 to predict 7-day risk of CLABSI. Competing events are discharge and death. Static models at catheter onset included logistic, multinomial logistic, Cox, cause-specific hazard, and Fine-Gray regression. Dynamic models updated predictions daily up to 30 days after catheter onset (i.e. landmarks 0 to 30 days), and included landmark supermodel extensions of the static models, separate Fine-Gray models per landmark time, and regularized multi-task learning (RMTL). Model performance was assessed using 100 random 2:1 train-test splits.

**Results** The Cox model performed worst of all static models in terms of area under the receiver operating characteristic curve (AUC) and calibration. Dynamic landmark supermodels reached peak AUCs between 0.741-0.747 at landmark 5. The Cox landmark supermodel had the worst AUCs (≤0.731) and calibration up to landmark 7. Separate Fine-Gray models per landmark performed worst for later landmarks, when the number of patients at risk was low.

**Discussion and Conclusion** Categorical and time-to-event approaches had similar performance in the static and dynamic settings, except Cox models. Ignoring competing risks caused problems for risk prediction in the time-to-event framework (Cox), but not in the categorical framework (logistic regression).

*Keywords:* risk prediction; central line-associated bloodstream infection; dynamic model; logistic regression; survival analysis

# 1. INTRODUCTION

Prediction models play an important role in providing decision support through individualized estimates of disease risk. Electronic health records (EHRs) are commonly used to develop prediction models for in-hospital outcomes. EHR data are complex longitudinal datasets that contain a large number of variables measured at irregular intervals. In the context of in-hospital outcomes, there is often no censoring in the sense that hospitals register data continuously from the time patients are admitted until they are discharged or deceased, and no patients are lost to follow-up between admission and discharge or death. EHR data are also characterized by the presence of competing events for many prognostic outcomes, i.e. there are competing events that preclude the event of interest from occurring, such as death.

When predicting prognostic outcomes in data without censoring but with competing events, several modeling approaches for categorical and time-to-event outcomes are possible. Operationalizing the outcome as a binary one (experiencing versus not experiencing the event of interest within time *t*) using logistic regression is common[1, 2]. Alternatively, modeling the time to the event of interest using standard Cox proportional hazards regression is often used[1, 2]. Although common, this does not explicitly address competing events and may therefore be expected to perform suboptimally[3, 4]. To account for competing events, time-to-event methods such as cause-specific hazard or Fine-Gray subdistribution hazard models can be used[5]. In the absence of censoring, multinomial regression can also model competing events as outcome categories in addition to the event of interest[6].

For outcomes during admission, there is often an interest in updating predictions over time as the health of the patient changes during their stay. For such dynamic prediction modeling, van Houwelingen described the landmarking approach[7]. Landmarking involves fitting a model that can make or update predictions at a series of time points during follow-up, known as landmarks (LMs), as a function of predictors measured up to the landmark time. This approach is straightforward to implement and computationally simple[8]. An alternative approach that was recently introduced is regularized multi-task learning (RMTL), which aims to jointly optimize predictions for multiple related tasks, such as predicting at multiple landmarks[9].

In this study, we aimed to compare various categorical and time-to-event approaches in developing both static and dynamic models, focusing on central line-associated bloodstream infections (CLABSI). CLABSI, occurring at least 48 hours after admission in the absence of infection at another site[10], is a priority for prevention due to its association with prolonged hospital stays, increased healthcare costs and elevated morbidity and mortality[11, 12, 13]. Our analysis utilized EHR data from University Hospitals Leuven (UZ Leuven) to predict the 7-day risk of CLABSI among hospitalized patients with catheters.

# 2. MATERIALS AND METHODS

## 2.1 Study design and participants

This is a retrospective cohort study of 27,478 patient admissions from the University Hospitals Leuven who were admitted to the hospital and received any type of central venous catheter between January 1$^{st}$, 2012 and December 31$^{st}$, 2013. The study sample included 30,862 patient-catheter episodes. A patient may have multiple catheter episodes if the patient received multiple catheters with more than 48 hours in between catheters. Details can be found in supplementary file 1.

## 2.2 Study outcome

Our primary focus is on the outcome of CLABSI, which is defined as any laboratory-confirmed bloodstream infection occurring in a patient with a central line or within 48 hours after its removal, in accordance with the 2019 definition published by Sciensano, the Scientific Institute of Public Health in Belgium[10]. This definition specifically excludes infections present on admission, secondary infections, skin contamination, and laboratory-confirmed bloodstream infection resulting from mucosal barrier injury. Further details can be found in supplementary file 1. The outcome to be predicted is CLABSI occurrence within 7 days of the moment of prediction (i.e. at baseline or at a later landmark). The time horizon of 7 days was chosen because this was considered the clinically meaningful time interval to intervene according to clinical experts, which included an infection preventionist, an intensivist, and a specialist from the central-line placement team. We discerned two competing events: (1) death or start of palliative care (predictions in palliative care are not actionable due to reduced monitoring intensity), and (2) discharge from hospital or catheter removal for more than 48 hours without CLABSI.

Among 30,862 catheter-episodes, 404 resulted in CLABSI within the first 7 days after catheter onset (1.31%). An additional 566 catheter-episodes encountered CLABSI beyond the initial 7 days following the onset of catheter episodes (1.83%). Overall, 970 catheter episodes resulted in CLABSI (3.14%) throughout the entire admission. Figure 1 showed the frequencies of the outcomes within 7 days across the catheter episodes at risks in each of the landmark subsets (LM $\leq$ 30).

## 2.3 EHR data and predictors

EHR data were extracted from various electronic sources including demographics information, patient admissions, discharges and transitions (e.g., transfers to intensive care units), catheter-related observations, patient medication prescriptions, comorbidities, laboratory tests and vital signs.

We used 21 predictors (Table 1), which were recorded routinely and chosen based on the domain knowledge collected from the clinical experts and from a systematic review[2]. Twenty of these were time-dependent variables.

## 2.4 Prediction models

### 2.4.1 Static models at the onset of catheter episodes

For static model, we intended to make prediction at the onset of catheter episodes, which occurred either upon catheter placement or upon the registration of the first catheter observation during the admission. Static models at catheter onset included logistic, multinomial logistic, Cox, cause-specific hazard, and Fine-Gray regression. Table S3 (supplementary file 3) provided a summary of the outcomes and the corresponding formulas for the static models[14, 15, 16, 17]. All above mentioned time-to-event models rely on the validity of the proportional hazards assumption. A potential solution to the robustness problem is to apply administrative censoring[18]. All observations are then censored at the target prediction horizon, ensuring that only outcome data directly associated to the survival probability within the specified time window of interest is used[18]. We fitted time-to-event models with and without administrative censoring. Notably, the Cox models, which ignore other competing risks are expected to overestimate the risk of the event of interest and be miscalibrated.

### 2.4.2 Dynamic models

For dynamic models, time-varying predictor information is used (Table 1).

The landmark approach in survival analysis involves selecting specific time points, known as "landmarks" $\{s_0, ..., s_L\}$ at which the risk estimates for an event of interest are updated, using the information on the individuals who survive up to that given landmark time point.

Let $w$ be the prediction window of interest. We aimed to create a model to estimate risk at landmark time $s$, knowing an individual's covariate values at $s$, namely $Z(s)$, conditioning on being event-free at $s$. To create the landmark model, landmark datasets are created for each landmark $s$, using only the data of individuals still at risk at $s$, and applying administrative censoring at $s+w$ to these individuals. Separate models can be fitted at each landmark $s$ for which a prediction is required. However, this is less practical and difficult to communicate with clinical users as a different prediction equation is used for each prediction time point of interest[8]. Alternatively, a landmark supermodel can be fitted after stacking landmark datasets into a super dataset (details can be found in supplementary file 2).

Landmark supermodels were fitted for binary logistic regression, multinomial logistic regression, Cox survival, cause-specific survival and Fine-Gray survival models. Table S4 (supplementary file 3) summarizes the corresponding formulas for predicting CLABSI with dynamic landmark models [19, 20, 21]. The concern in developing a landmark supermodel based on the Fine-Gray approach was in constructing landmark datasets that properly accounted for competing events that happened before the landmark in the setting of subdistribution hazard[21]. Liu et al. extended the landmark method to the Fine-Gray model and proposed the landmark proportional subdistribution hazards (PSH) supermodel by transforming each landmark subset into the counting process style before stacking all landmark datasets (details can be found in supplementary file 2)[22]. Additionally, time-varying inverse probability censored weighting (IPCW) needed to be calculated for the subjects who experienced competing risks[23]. With these necessary changes, the Fine-Gray supermodel stands out from the other supermodels. For comparison, we also constructed separate Fine-Gray models for each landmark.

Lastly, we implemented regularized multi-task learning (RMTL) for simultaneous learning of distinct logistic regression tasks[24]. We defined tasks as landmark specific models, with t=31 tasks in total. Generally, the algorithm uses gradient descent to estimate the task-specific coefficients by minimizing the summation of logistic loss functions across all tasks and two regularization terms, parameterized by $\lambda_1$ for cross-task regularization (knowledge transfer across tasks), and $\lambda_2$ for the L2-norm or ridge regularization. For cross-task regularization, we used network-based relatedness to incorporate time-smoothness in the estimated coefficients, which has the effect of shrinking the coefficients of adjacent landmarks towards each other. We tuned $\lambda_1$ (for knowledge transfer across landmarks) using 5-fold cross-validation on the training set and set the $\lambda_2$ hyperparameter to 0 (no regularization towards zero). The variables in each task train set have been standardized by subtracting the mean of the variable and dividing by the standard deviation. The task test sets have been standardized using the mean and standard deviation of each variable from the corresponding task train set.

## 2.5 Statistical analysis

### 2.5.1 Model building and validation

Table 2 lists all models that were fitted. All static and dynamic models used the same 21 predictors, but static models only used predictor values at the onset of catheter episodes. Systolic blood pressure was modeled using restricted cubic splines with three knots. White blood cell count and C-reactive protein were log-transformed due to their right-skewed distributions. For dynamic models, smooth baseline hazards were assumed, by including the linear and quadratic landmark time variables over the stacked landmark datasets in model-fitting. In addition, we tested the landmark-covariate interactions for each covariate via the Wald test (α=0.05). We found that the effects of being on an ICU unit was significantly dependent on the landmark time (linear and quadratic), and included the interaction terms in our dynamic models to capture the time-dependent effect of this predictor. We used repeated data splitting to develop and validate the models (Figure 2). The procedure is as follows:
- A random sample of two thirds of hospital admissions from the landmark dataset was used for training, the remaining one third of the admissions were used for model validation. For dynamic models, all landmarks and all catheter episodes of one hospital admission either fell completely in training or test data.

- The candidate models were fitted on the training data.
- The fitted models were used to obtain predicted risk of CLABSI for the test data.
- Model performance measures were evaluated in the test data.
- The steps above were repeated 100 times.

We assessed model performance on test data in terms of discrimination using the area under the receiver operating characteristic curve (AUC)[25], calibration (measuring how well the estimated probabilities match the observed probabilities[26]) and overall performance using the scaled Brier score[25] (Table 3). As there was no censoring in the dataset, all performance measures were assessed by treating the outcome as binary. To do so, we evaluated the estimated probability of CLABSI within 7 days against the occurrence of CLABSI within 7 days (yes vs no).

For dynamic models, landmarks beyond day 30 were not considered due to the limited number of catheter episodes (872 catheter episodes, 2.83%) remaining at risk after day 30.

### 2.5.2 Missing value imputation

Missingness percentages of the predictors at all landmarks were shown in Table S5 (supplementary file 4). One variable (admission source) was imputed with mode value due to its low percentage of missingness (less than 3%). For the remaining variables with missing values, missing data imputations have been performed on each training set using an adaptation of the missForest algorithm for prediction settings[31]. The missing values were first imputed with mean/mode and then iteratively imputed using random forest models for 5 iterations. The outcome was not included in the imputation and test datasets were imputed using the missForestPredict imputation models learned on the matching training dataset to mimic prediction model use at the bedside[32].

### 2.5.3 Sample size and software

The sample size calculation was detailed in supplementary file 5. Sample size was calculated using the *pmsampsize* package[33]. All analyses were performed using R v4.3.2. Software packages used for building baseline and dynamic models were shown also in Table 2. Details regarding the R code for fitting and evaluating the models can be accessed via supplementary file 9. The codes were illustrated using the UZ Leuven EHR data. As it is not permitted to share the original data, a manually created example dataset is shared in supplementary file 2 with sensitive information being replaced.

## 2.6 Ethics

The study adhered to the principles of the Declaration of Helsinki (current version), the principles of Good Clinical Practice (GCP), and all relevant regulatory requirements. Ethical review was sought from the Research Ethics Committee UZ/KU Leuven, Belgium, and local ethics committees at participating hospitals. The collection, processing and disclosure of personal data, such as patient health and medical information were in compliance with applicable personal data protection and the processing of personal data (Directive 95/46/EC and Belgian law of December 8, 1992 on the Protection of the Privacy in relation to the Processing of Personal Data). Patient stay identifiers were coded using the pseudo-identifier available in the data warehouse of the participating hospital.

# 3. RESULTS

### 3.1 Static models at the onset of catheter episodes

All models showed similar discrimination and calibration performances except Cox proportional hazard models (Table 4, Supplementary file 6). Cox models had lower AUC values and clear miscalibration compared to the other models. The cause-specific model without administrative censoring had the highest AUC (median 0.721). In terms of calibration, the cause-specific and Fine-Gray model with administrative censoring, as well as binary and multinomial logistic models showed well-calibrated results. For overall performance, cause-specific models, either with or without administrative censoring, as well as multinomial logistic model showed higher scaled Brier scores than the other models.

### 3.2 Dynamic models

Generally, discrimination increased from landmark day 0 to landmark day 5, then slightly decreased up to landmark day 16, and decreased strongly thereafter (Figure 3). It can be noticed that the performance metrics of separate Fine-Gray landmark model started to diverge from others since landmark day 14, coinciding with the emergence of convergence issues from landmark day 14 onward (details can be found in supplementary file 7). The Cox landmark supermodel had clearly lower AUC than all other models up to landmark 7. The landmark cause-specific supermodel had the highest AUC (mean AUC up to 0.739) for predictions up to 3 days after the onset of catheter episodes. For predictions between 4 and 13 days after placement, the landmark multinomial logistic model had a slightly higher AUC than other models (mean AUC up to 0.747). After landmark day 13, the landmark logistic and Fine-Gray model had higher AUC values than the other models until landmark day 20. The approach based on separate Fine-Gray models per landmark had worse AUCs than other models from landmark 10 onwards.

The Cox landmark supermodel was less competitive in terms of calibration compared to other dynamic models. The other models demonstrated comparable performance. For binary and multinomial logistic landmark supermodels, calibration deteriorated after landmark day 14 in terms of O/E ratio. In contrast, the Fine-Gray and cause-specific landmark supermodels maintained stable O/E ratios. For all supermodels and RMTL, calibration slopes were above 1 before landmark day 7, implying that risk estimates tended to be too close to the average CLABSI risk. For separate Fine-Gray landmark models, calibration slopes were always below 1, indicating overfitting.

In terms of scaled brier score, the multinomial logistic landmark supermodel was superior to the other models before landmark day 11. Afterwards, the binary logistic landmark supermodel showed a slight advantage until landmark day 18.

## 4. DISCUSSION

In this study we compared different statistical methods for predicting the risk of CLABSI within 7 days during hospital admission using EHR data in which there was no censoring. Both static and dynamic models were fitted and compared. Discrimination, calibration and overall predictive performance showed relatively minor distinctions, except the Cox proportional hazards models which performed noticeably worse than the other models. This was expected as the Cox model ignores competing risks and resulted in overestimation of risks[34]. Logistic regression models performed very well, as in contrast to Cox models, competing events were not censored (and assumed to have the same CLABSI risk as those still in the risk set) but counted as non-events. Multinomial regression with competing events as separate outcome classes had a small advantage over binary logistic regression, possibly attributed to its ability to preserve essential information concerning the different outcome categories. The performances of RMTL model were moderately better than that of Cox and separate Fine-Gray landmark models, without clear evidence indicating their superiority over the other models.

In a recent systematic review covering 16 prediction models for CLABSI, 13 models used a binary outcome (9 with logistic regression, 2 with XGBoost, 1 with random forests, and 1 with naïve Bayes) and 3 modeled time to CLABSI using Cox proportional hazards regression[2]. None of the models considered competing events. Unfortunately, only one model fixed the prediction horizon[35], the other 15 models considered CLABSI at any time during admission. In our study, we fixed the time horizon to ensure a clear and interpretable definition of the outcome. Additionally, in the systematic review, only one model attempted dynamic prediction and 15 models focused on static prediction at the moment of catheter onset. Unfortunately, all models had a high risk of bias. Based on this systematic review[2], Cox models remained the second most commonly used method for CLABSI prediction. Researchers should take note of the potential overestimation and loss of discrimination associated with the Cox model observed in the current study and previous research[34].

It is worth noticing that in this study we use the landmark approach in the competing risk setting, applying it to the cause-specific and Fine-Gray model. Though there have been studies that applied a landmark approach to the Cox proportional hazards model[36, 37], there are few studies about its application in competing risks survival models. Nicolaie et al. extended the landmark model for ordinary survival data to address the problem of dynamic prediction in the presence of competing risks, based on the cause-specific hazards and dynamic pseudo-observation, respectively[21, 38]. In this study, we fitted the landmark supermodel to the stacked dataset across all landmarks by smoothing baseline hazards[8]. We also fitted a separate Fine-Gray model per landmark. However, these models had worse performance than the Fine-Gray landmark supermodel[22, 23]. This might be attributed to the substantial decrease in sample size at later landmarks, in combination with the fact that separate models do not borrow information from adjacent landmarks. While separate landmark models can be developed for all approaches, our focus here is solely on the Fine-Gray approach for simplicity.

Out study also has limitations. To simplify the problem, we made assumptions on the linear and quadratic landmark time effects as well as their interactions with ICU in landmark supermodels. These assumptions may be too strong, but relaxing the assumptions may require a lot more data. Developing one single landmark supermodel made it more vulnerable to misspecification of the model, considering that unexpected changes may happen over time in the relationship between predictors and hazard[39].

As a next step, it is worthwhile to also deploy machine learning algorithms, which may automatically detect and include nonlinear and nonadditive associations between predictors and outcome. This is further discussed in another paper, which specifically delved into random forest implementations of similar modeling approaches for the same application [40].

## 5. CONCLUSIONS

In conclusion, our study compared various statistical approaches for predicting CLABSI within 7 days during hospital admission using EHR data without censoring from UZ Leuven. In the absence of censoring, time-to-event, logistic and multinomial regression models yielded comparable predictive performance in static and dynamic prediction. However, Cox models, which overlooked competing events that may preclude the occurrence of the outcome of interest, exhibited inferior performance. Our study applied the landmark approach to competing risk settings, highlighting the importance of considering competing events in predictive modeling. Overall, our findings underscored the significance of appropriate model selection and the consideration of competing risks for accurate risk prediction in clinical settings.

# Acknowledgements
The funding sources had no role in the conception, design, data collection, analysis, or reporting of this study.

# Author contribution
**GAO Shan**: Conceptualization, Data curation, Formal analysis, Investigation, Methodology, Software, Visualization, Writing - original draft. **ALBU Elena**: Data curation, Investigation, Formal analysis, Software, Writing - review & editing. **PUTTER Hein**: Methodology, Writing - review & editing. **Stijnen Pieter**: Conceptualization, Data curation, Funding acquisition, Writing - review & editing. **RADEMAKERS Frank**: Supervision, Writing - review & editing. **COSSEY Veerle**: Supervision, Writing - review & editing. **DEBAVEYE Yves**: Supervision, Writing - review & editing. **JANSSENS Christel**: Supervision, Writing - review & editing. **VAN CALSTER Ben**: Conceptualization, Funding acquisition, Methodology, Project administration, Supervision, Writing - review & editing. **WYNANTS Laure**: Conceptualization, Funding acquisition, Methodology, Project administration, Supervision, Writing - review & editing

# Funding
This work was supported by the Internal Funds KU Leuven [grant C24M/20/064].

# Conflicts of interest
The authors declare that they have no conflicts of interests to disclose.

# Data availability
The data underlying this article cannot be shared publicly due to for the privacy of individuals that participated in the study.

Table 1: Covariates from the UZ Leuven EHR dataset used in the prediction models

| Variable Category | Variables | Description | Further Information |
| --- | --- | --- | --- |
| **catheter types** | Central venous catheter | No (0), Yes (1) | In the last 24h |
| | Port-a-cath | No (0), Yes (1) | In the last 24h |
| | Tunneled central venous catheter | No (0), Yes (1) | In the last 24h |
| | Peripherally inserted central catheter | No (0), Yes (1) | In the last 24h |
| **catheter location** | Subclavian | No (0), Yes (1) | In the last 24h |
| | Jugular | No (0), Yes (1) | In the last 24h |
| **medication** | Total parenteral nutrition | No (0), Yes (1) | In the last 7 days |
| | Antibacterials for systematic use | No (0), Yes (1) | In the last 7 days |
| | Antineoplastic agents | No (0), Yes (1) | In the last 7 days |
| **CLABSI history** | History of CLABSI | No (0), Yes (1) | In the last 3 months |
| **comorbidity** | Tumor | No (0), Yes (1) | Before current LM |
| | Lymphoma | No (0), Yes (1) | Before current LM |
| | Transplant | No (0), Yes (1) | Before current LM |
| **physical ward** | ICU unit | No (0), Yes (1) | In the last 24h |
| **care modules** | Mechanical ventilation | No (0), Yes (1) | In the last 24h |
| | Temperature | Unit: °C | Maximum value in the last 24h |
| | Systolic blood pressure | Unit: mmHg | Last value in the last 24h |
| **laboratory test** | WBC count | Unit: $10^9$/L | Last value in the last 24h |
| | CRP | Unit: mg/L | Last value in the last 24h |
| | Positive culture, of any other type than blood | No (0), Yes (1) | In the last 17 days |
| **baseline variable** | Whether patients admitted from: home | No (0), Yes (1) | |

All are time-varying except the baseline variable. Due to the collections of various data source are at irregular time interval, we take the dynamic variable values per landmark (24h). Details can be found in supplementary file 2.

LM: landmark; ICU: intensive care unit; WBC: white blood count; CRP: C-reactive protein

Table 2: Summary of statistical methods used for baseline and dynamic models

| Model name | Regression model | Competing risks | Administrative censoring | R package and function |
| --- | --- | --- | --- | --- |
| **Static** | | | | |
| Cox | Cox | No | No | survival::coxph |
| Cox-ac | Cox | No | Yes | survival::coxph |
| CS | Cause-specific | Yes | No | riskRegression::CSC |
| CS-ac | Cause-specific | Yes | Yes | riskRegression::CSC |
| FG | Fine-Gray | Yes | No | riskRegression::FGR |
| FG-ac | Fine-Gray | Yes | Yes | riskRegression::FGR |
| LR | Logistic | No | NA | stats::glm |
| MLR | Multinomial Logistic | Yes | NA | nnet:::multinom |
| **Dynamic** | | | | |
| LM Cox | Cox | No | Yes | survival::coxph |
| LM CS | Cause-specific | Yes | Yes | riskRegression::CSC |
| LM FG | Fine-Gray | Yes | Yes | survival::coxph; survival::finegray |
| FG-sep | Fine-Gray | Yes | Yes | riskRegression::FGR |
| LM LR | Logistic | No | NA | stats::glm |
| LM MLR | Multinomial Logistic | Yes | NA | nnet:::multinom |
| RMTL-ts | RMTL | No | NA | RMTL::MTL |

NA: not applicable.

Table 3: Assessment of predictive performance

|  | Explanation |
|---|---|
| **Discrimination** | |
| AUC | Ranges in value from 0 to 1. |
| **Calibration** | |
| Calibration slope | Target value is 1. A slope < 1 suggests that estimated risks are too extreme, i.e., too high for patients who are at high risk and too low for patients who are at low risk. A slope > 1 suggests the opposite, i.e., that risk estimates are too moderate[27]. |
| O/E ratio | Target value is 1. Ratio of overall observed outcome proportion to average estimated risk. The O/E ratio < 1 suggests that the model tends to overestimate the risk[28]. |
| ECI | Target value is 0. An averaged squared difference of the predicted probabilities with the estimated observed probabilities[29]. The larger the ECI, the more uncalibrated the model. |
| **Overall** | |
| Scaled Brier | Ranges in value from 0 to 1. The percentage reduction in Brier score compared to a null model. The higher the scaled brier score is, the better the predictions are discriminated and calibrated[30]. |

AUC: Area under the receiver operating characteristic (ROC) curve; O/E ratio: observed rate/expected rate; ECI: estimated calibration index.

Table 4: Summary of the performance measures (mean with 95% CI) for static models

| Model | AUC | Calibration Slope | O/E ratio | ECI | Scaled Brier |
|---|---|---|---|---|---|
| **Cox-ac** | 0.649 (0.644, 0.654) | 1.083 (1.052, 1.115) | 0.603 (0.592, 0.615) | 0.017 (0.016, 0.019) | -0.001 (-0.002, -0.000) |
| **Cox** | 0.656 (0.650, 0.662) | 1.416 (1.373, 1.459) | 0.583 (0.572, 0.593) | 0.014 (0.013, 0.014) | 0.001 (-0.000, 0.001) |
| **CS-ac** | 0.715 (0.711, 0.718) | 0.940 (0.920, 0.960) | 1.028 (1.009, 1.047) | 0.005 (0.004, 0.005) | 0.008 (0.008, 0.009) |
| **CS** | 0.721 (0.718, 0.725) | 1.204 (1.180, 1.228) | 1.052 (1.033, 1.071) | 0.003 (0.002, 0.003) | 0.009 (0.009, 0.010) |
| **FG-ac** | 0.705 (0.702, 0.709) | 0.913 (0.893, 0.932) | 1.011 (0.992, 1.030) | 0.005 (0.005, 0.006) | 0.007 (0.006, 0.008) |
| **FG** | 0.718 (0.715, 0.722) | 0.870 (0.854, 0.885) | 1.014 (0.995, 1.032) | 0.005 (0.005, 0.006) | 0.005 (0.004, 0.006) |
| **LR** | 0.706 (0.702, 0.709) | 0.909 (0.890, 0.929) | 1.009 (0.990, 1.028) | 0.005 (0.005, 0.006) | 0.007 (0.006, 0.008) |
| **MLR** | 0.713 (0.710, 0.717) | 0.938 (0.917, 0.958) | 1.009 (0.990, 1.028) | 0.004 (0.004, 0.005) | 0.009 (0.008, 0.009) |

Figure 1: Frequency of outcomes within 7 days for each of the landmark subsets (LM ≤ 30). The total height of the bar is the number at risk.

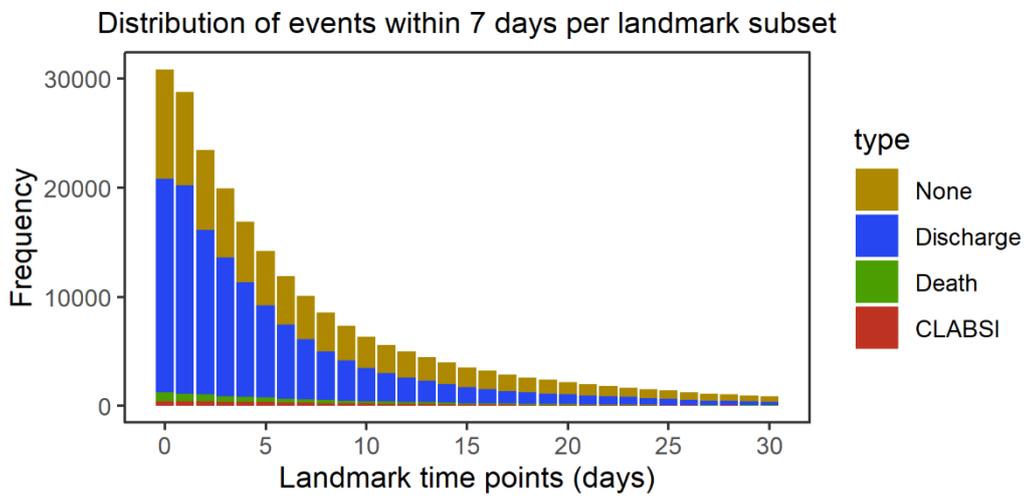

Figure 2: Creation of train-test data for estimating performance.

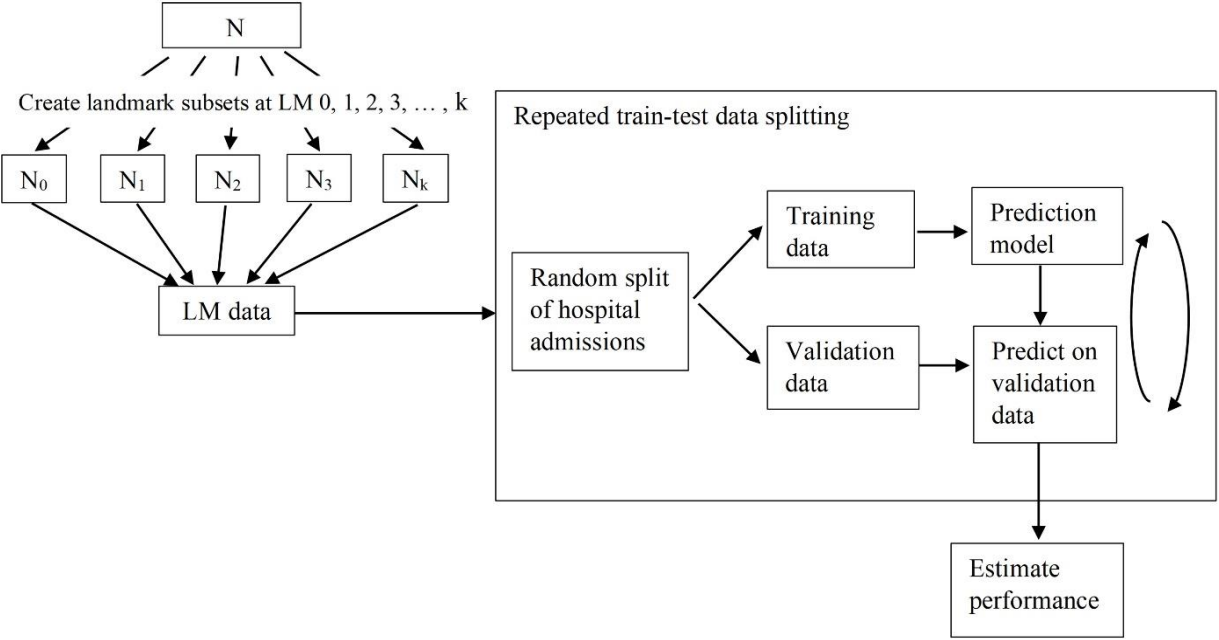

Figure 3: Comparison of performance metrics of dynamic models across landmarks. The vertical Y axis was truncated for clarity. Minimum mean observed AUC was 0.535, minimum /maximum mean observed calibration slope was 0.093/ 1.742, maximum mean observed ECI was 0.386, minimum mean observed scaled BS was -0.133.

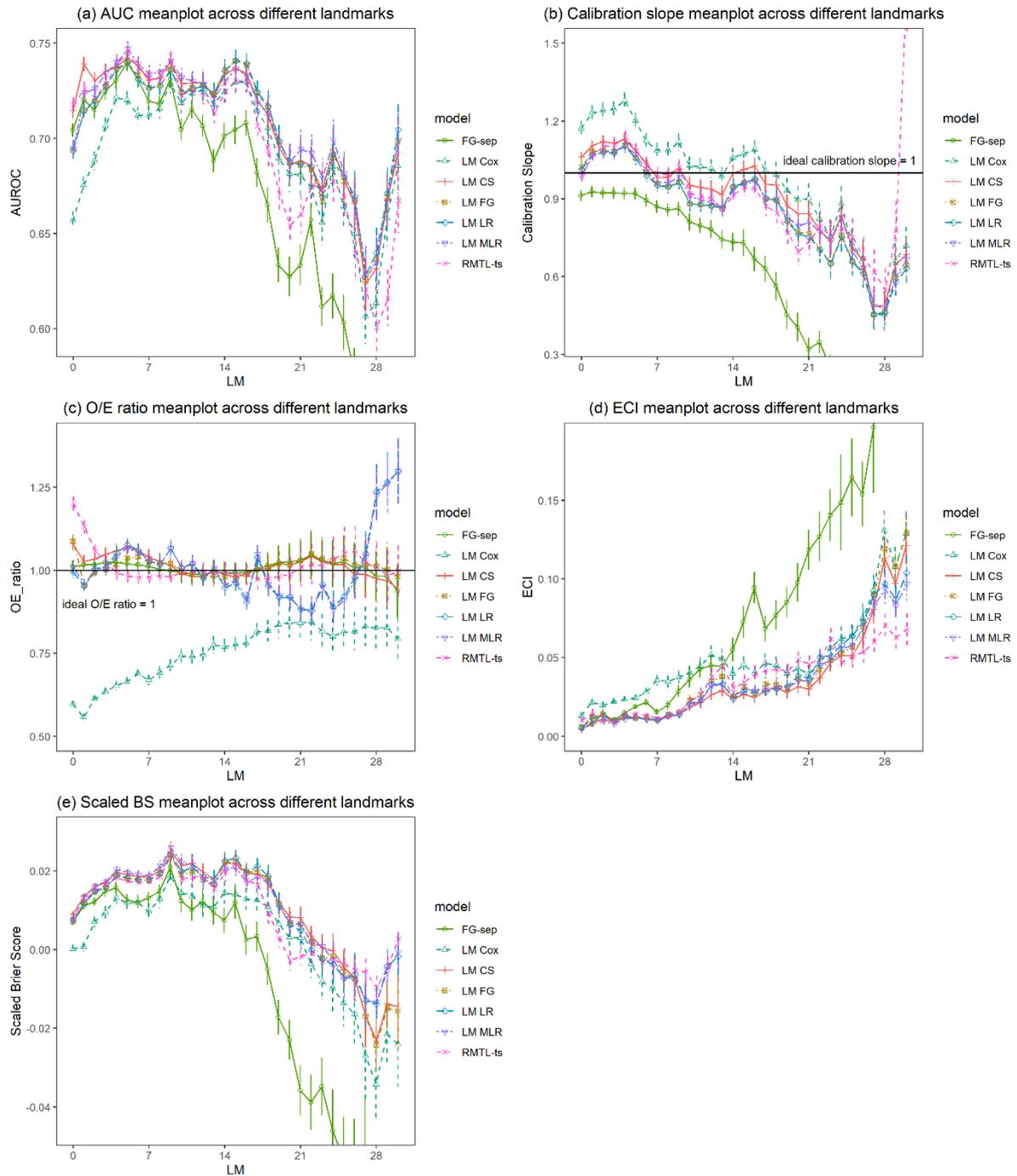

# Supplementary file 1: CLABSI in UZ LEUVEN

The retrospective cohort study consists of patients from the University Hospitals Leuven who were admitted to the hospital and received a catheter between January 2012 and December 2013.

- **Inclusion criteria**

Only hospital stays in participating hospitals with admission starting from January 1st 2012 up to December 30th 2013 were analyzed. Only patients who had a catheter of the following types were included, in accordance with the definitions of CLABSI of the Hospital Hygiene Department:

- Deep venous catheter
- Peripherally inserted central catheter (PICC) (open and valve)
- Midline catheter (open and valve)
- Tunneled dialysis catheter
- Non-tunneled dialysis catheter
- Umbilical catheter
- Port-a-cath
- Hickman catheter

Excluded were: rapid infusion system (RIS) catheters, Swan-Ganz catheters, Coolgard catheter introducers, pacemakers, arterial catheter, peripheral venous catheter, Extracorporeal membrane oxygenation (ECMO), intra-aortic balloon pump (IABP) Patient-controlled epidural analgesia (PCEA), Pulse index Contour Continuous Cardiac Output (PICCO) .

Patients that had only a dialysis catheter were included only if they have an ICU admission between January 2012 and December 2013, due to the data extraction constraint.

- **Exclusion criteria**

Patients in the neonatology department were documented using a paper-based workflow before October 2013 and did not have electronic records in the system. Thus hospital admissions for patients under the age of 12 weeks have been excluded from the analysis.

- **Catheter episodes**

Considering that there were situations that patients received more than one catheter simultaneously or consecutively in ICU or in other hospital wards, it is difficult to distinguish the effect of those catheters on the risk of CLABSI. Thus, a scheme was developed to make a difference between these situations.

- When the patient received only one catheter, this was regarded as one observation. Its time at risk is the time interval from the catheter placement to the catheter removal plus 48 hours, according to the definition of CLABSI [1].

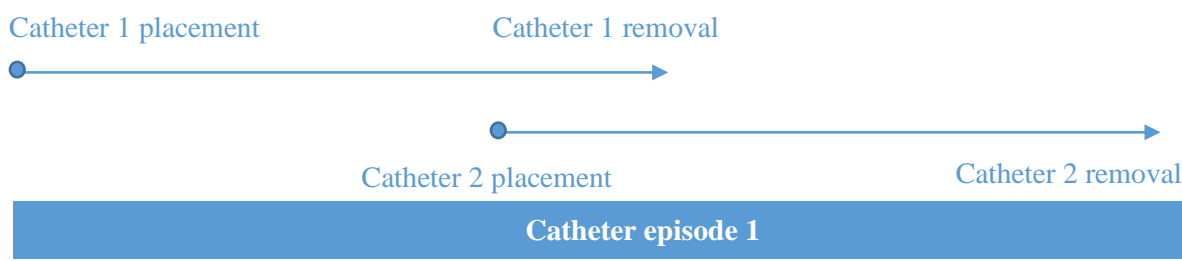

- If the same patient received two catheters and the time interval between these two catheters was less than 48 hours, we treat them as one observation, which means that their time at risk are counted together, from the start of the earlier catheter to the end of the later catheter.

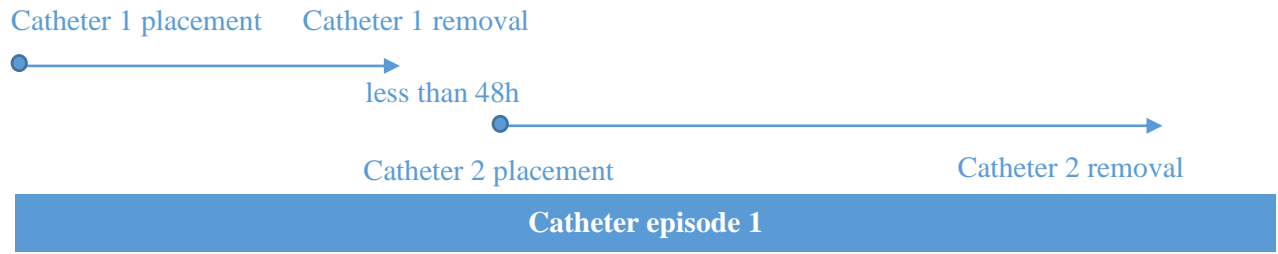

- If the same patients received two catheters and the time interval between these two catheters was more than 48 hours, we treat them as two observations, that is, their time at risk are counted separately.

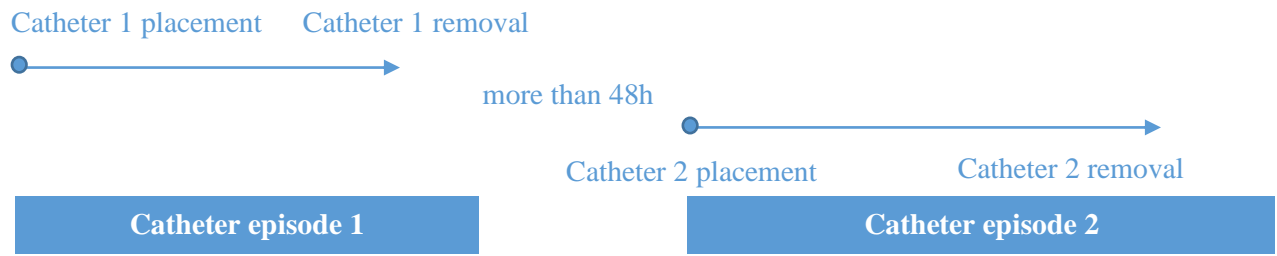

- **Outcome**

There are three types of events (CLABSI, death and discharge) considered in this analysis.

- CLABSI: any laboratory-confirmed bloodstream infection (LCBSI) for a patient with central line or within 48 hours after the central line removal. The CLABSI definition follows the Sciensano definition published in 2019 [2], with specific criteria excluding infection present on admission, secondary infections, skin contamination, and mucosal barrier injury LCBSI. Symptoms criteria are not checked, and it is assumed that cultures are ordered based on relevant symptoms. The time window for secondary infections is not clearly defined in Sciensano, and we use the time window of the 17 days, considering the CLABSI episode length of 14 days plus the time window of 3 days.
- Discharge: hospital discharge or 48 hours after catheter removal, whichever happens first. A patient is still considered at risk within 48 hours after catheter removal.
- Death: Either the first contact with palliative care during admission, transfer to palliative care or patient death, whichever happens first. Patients are not closely monitored in palliative care and predictions for this ward are deemed non-actionable, thus of limited value since there are minimal opportunities to prevent CLABSI events in this context.

# Supplementary file 2: stacked dataset

The landmarking approach for dynamic prediction of survival was initially described by van Houwelingen [1]. In brief, at a given landmark time $s$ where a prediction is to be made, the data are restricted to individuals who have not yet experienced the event. Predictor values available up to the landmark time are used as covariates in a model for the probability of survival up to some time horizon, conditional on survival to the landmark time. Typically, the focus is on survival to a single time horizon $w$, and censoring is imposed at $w$ so that only events up to that time $s+w$ are used in the survival analysis. In this approach, the dataset is transformed into multiple censored datasets based on predefined $s$ and $w$. Traditionally, a separate cox proportional hazards model can be applied to each landmark dataset and predictions can then be made at each landmark time point. Moreover, a supermodel can be fitted on the stacked super dataset and the landmark supermodel may combine these models by introducing smoothing to permit risk prediction at any landmark. Then dynamic risk prediction can be performed by using the most up-to-date value of patients' covariate values.

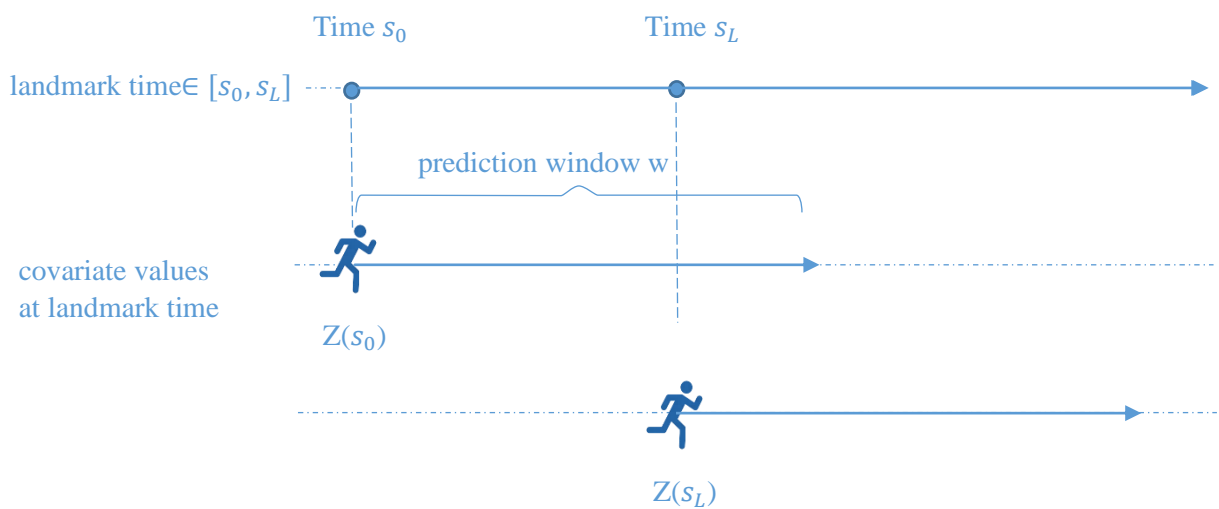

To fit a landmark Cox supermodel [3,4], a stacked dataset is constructed by: (i) firstly selecting a set of landmark points $s$ from $[s_0, s_L]$; (ii) then creating a landmark subset by selecting the subjects who have not yet failed from any cause at $s$ and adding an administrative censoring at the prediction horizon $s+w$; (iii) finally stacking all the individual landmark subsets into a super prediction dataset.

The following table is an example of the stacked super dataset (with pseudo-anonymization) which is used to develop the dynamic models. The original data is not allowed to share, thus sensitive information are replaced with manually created synthetic data in the following table.

Table S1: example data for landmark supermodels (except Fine-Gray supermodel)

| ID | LM | eventtime | type[a] | ICU unit | CRP |
|---|---|---|---|---|---|
| 1 | 0 | 4.42 | 1 | 0 | 28.6 |
| 1 | 1 | 4.42 | 1 | 0 | 50.7 |
| 1 | 2 | 4.42 | 1 | 0 | 46.2 |
| 1 | 3 | 4.42 | 1 | 1 | 46.7 |
| 1 | 4 | 4.42 | 1 | 0 | 21.2 |
| 2 | 0 | 7.00[b] | 0 | 1 | 86.1 |
| 2 | 1 | 8.00[b] | 0 | 1 | 99.7 |
| 2 | 2 | 9.00[b] | 0 | 1 | 87.5 |
| 2 | 3 | 9.34 | 1 | 1 | 51.2 |
| 2 | 4 | 9.34 | 1 | 1 | 40.7 |
| 2 | 5 | 9.34 | 1 | 1 | 27.5 |
| 2 | 6 | 9.34 | 1 | 1 | 19.6 |
| 2 | 7 | 9.34 | 1 | 1 | 29.3 |
| 2 | 8 | 9.34 | 1 | 1 | 17.8 |
| 2 | 9 | 9.34 | 1 | 1 | 9.6 |
| 3 | 0 | 1.29 | 2 | 1 | 90.1 |
| 3 | 1 | 1.29 | 2 | 1 | 131.4 |
| 4 | 0 | 4.56 | 3 | 0 | 41.1 |
| 4 | 1 | 4.56 | 3 | 0 | 157.3 |
| 4 | 2 | 4.56 | 3 | 0 | 167.9 |
| 4 | 3 | 4.56 | 3 | 0 | 134.5 |
| 4 | 4 | 4.56 | 3 | 0 | 41.6 |

LM: landmark time; eventtime: time when any type of event happened; type: type of the event; ICU unit: binary indicator, whether the patient now (at the exact second of the current LM) is in ICU; CRP: continuous variable, last value of C-reactive protein test since previous LM. Unit: mg/L.

[a]type=1 (CLABSI); type=2 (Death); type=3 (Discharge); type=0 (Censored)

[b]Discharge here means catheter removal. As prediction time window is 7-day from each landmark time, individuals who are free of event up to 7 days of follow-up from the corresponding LM are administratively censored.

For landmark super subdistribution models, we used another expanded example dataset here for explanation. The standard estimator of the cause-specific cumulative incidence function can be written as a Kaplan-Meier type product-limit estimator with the estimator $d_j(t_i)/r(t_i)$ replaced by an estimator of the subdistribution hazard ($r(t_i)$ is the observed number at risk). This estimator of the subdistribution hazard $\hat{\lambda}_j(t_i)$ has the form $(d_j(t_i)/r^*(t_i))$ with $r^*(t_i)$ obtained by reweighting individuals who had a competing event before $t_i$. Thus, by creating a data set where individuals with an earlier competing event still contribute to the risk set with a weight, we can estimate the subdistribution hazard using the standard counting process approach. The following example is an expanded dataset created based on the synthetic table above [changes are marked in red].

It shall be noted that we create this expanded dataset by reweighting each landmark subset then combining them into the stacked set.

Table S2: example data for Fine-Gray landmark supermodel

| ID | Tstart[a] | Tstop[b] | status[c] | MS_is_ICU_unit | LAB_CRP_last | weight.cens | count | failcode |
|---|---|---|---|---|---|---|---|---|
| 1 | 0 | 4.42 | 1 | 0 | 28.6 | 1 | 1 | 1 |
| 1 | 1 | 4.42 | 1 | 0 | 50.7 | 1 | 1 | 1 |
| 1 | 2 | 4.42 | 1 | 0 | 46.2 | 1 | 1 | 1 |
| 1 | 3 | 4.42 | 1 | 1 | 46.7 | 1 | 1 | 1 |
| 1 | 4 | 4.42 | 1 | 0 | 21.2 | 1 | 1 | 1 |
| 2 | 0 | 7.00 | 0 | 1 | 86.1 | 1 | 1 | 1 |
| 2 | 1 | 8.00 | 0 | 1 | 99.7 | 1 | 1 | 1 |
| 2 | 2 | 9.00 | 0 | 1 | 87.5 | 1 | 1 | 1 |
| 2 | 3 | 9.34 | 1 | 1 | 51.2 | 1 | 1 | 1 |
| 2 | 4 | 9.34 | 1 | 1 | 40.7 | 1 | 1 | 1 |
| 2 | 5 | 9.34 | 1 | 1 | 27.5 | 1 | 1 | 1 |
| 2 | 6 | 9.34 | 1 | 1 | 19.6 | 1 | 1 | 1 |
| 2 | 7 | 9.34 | 1 | 1 | 29.3 | 1 | 1 | 1 |
| 2 | 8 | 9.34 | 1 | 1 | 17.8 | 1 | 1 | 1 |
| 2 | 9 | 9.34 | 1 | 1 | 9.6 | 1 | 1 | 1 |
| 3 | 0 | 1.29 | 2 | 1 | 90.1 | 1 | 1 | 1 |
| 3 | 1.29 | 4.42 | 2 | 1 | 90.1 | 1 | 2 | 1 |
| 3 | 1 | 1.29 | 2 | 1 | 131.4 | 1 | 1 | 1 |
| 3 | 1.29 | 4.42 | 2 | 1 | 131.4 | 1 | 2 | 1 |
| 4 | 0 | 4.56 | 3 | 0 | 41.1 | 1 | 1 | 1 |
| 4 | 1 | 4.56 | 3 | 0 | 157.3 | 1 | 1 | 1 |
| 4 | 2 | 4.56 | 3 | 0 | 167.9 | 1 | 1 | 1 |
| 4 | 3 | 4.56 | 3 | 0 | 134.5 | 1 | 1 | 1 |
| 4 | 4.56 | 9.34 | 3 | 0 | 134.5 | 1 | 2 | 1 |
| 4 | 4 | 4.56 | 3 | 0 | 41.6 | 1 | 1 | 1 |
| 4 | 4.56 | 9.34 | 3 | 0 | 41.6 | 1 | 2 | 1 |

[a]Tstart = LM

[b]Tstop = eventtime

[c]status = type

In the above table, events of CLABSI was observed at 4.42 for landmark time 0 to 4 and at 9.34 for landmark time 3 to 9. In the expanded table, the first two individuals didn't change. However, the individual 3, who had a competing event Death at 1.29, is spread over several rows (actually, they are the time when CLABSI happens: 4.42). Same happens on individual 4. As the time when competing event Discharge happened at a later time (4.56) than 4.42, thus it is only spread over the rows when a later CLABSI happened (9.34). Due to administrative censoring that we applied here, censoring time can only be 7+LM, thus the weight did not change and was always 1.

# Supplementary file 3: methodology
- *Baseline models*

Logistic regression is used to model the event, which is an indicator of whether CLABSI occurred within the 7 days after baseline but ignoring when the event happened exactly. We used our baseline datasets and related the probability of CLABSI occurring in an interval to a logistic function of the risk factors [5].

Let N be the number of individuals in the dataset, each with a set of k risk factors, $Z' = (z_1, z_2, ..., z_k)$, measured at the beginning (baseline) of follow-up period of length T. Defining a dichotomous response variable Y by

$$Y = \begin{cases} 1, & \text{if the event of interest happens during follow up;} \\ 0, & \text{otherwise;} \end{cases}$$

then the logistic function of risk factors $Z$ with coefficients $\beta_0$ and $\beta$ is given as follows:

$$P(Y = 1|Z) = \{1 + \exp[-\beta_0 - \beta'Z]\}^{-1} \quad (1)$$

The parameter $\beta_0$ is the intercept for the logistic model and the $Z$ represent the covariates that are available at baseline.

When there are competing events, these can be taken into account in the outcome indicator. Assuming J+1 possible events, multinomial logistic regression is a generalized linear model used to estimate the probabilities for each event j = 0, 1, ... , J (assuming the reference level is event j=0). It models the probability of the 1st category (always assuming the event of interest is 1st category) of a dependent variable Y, using a set of explanatory variables $Z$ [6]:

$$P(Y = 1|Z) = \frac{\exp(\beta_{01}+\beta'_1 Z)}{1+ \sum_{j=1}^{J} \exp(\beta_{0j}+\beta'_j Z)} \quad with\ j = 1,...,J \quad (2)$$

*where $\beta_j$* is the row vector of regression coefficients of $Z$ for the *j*th category of Y. The category 0 denotes the baseline category, which, in this case study, refers to those patients who have not experienced any type of event excluding CLABSI, death or discharge.

When the outcome is of a time-to-event nature, survival models are commonly advised. The basic approach is the Cox proportional hazards model. We denote by T the time between a baseline date and the date of an event, the corresponding survival function is given by

$$S(T|Z, \alpha) = \exp\left(-\int_0^T \lambda_0(t) \exp(\beta'Z)\, dt\right) \quad (3)$$

where $\lambda_0(t)$ denotes baseline hazard, which refers to the probability that a person with all zero values for covariates will experience the event in the next instant if the person survives to *t*. Note that the coefficients vector $\beta$ for the risk factors $Z$ are not necessarily the same as the coefficients $\beta'$ in (1).

Under competing risks setting, we denote by D ∈ {1, …, J} the different events. We assume that $\{D = 1\}$ is the event of interest. The absolute risk formula is given by [7]:

$$F_1(T|Z) = \int_0^T S(t|Z)\lambda_1^{cs}(t)dt \quad (4)$$

where $F_1(T|Z)$ and $\lambda_1^{cs}(t)$ denotes the cumulative incidence function (CIF) and the cause-specific hazard rate respectively for the target event of interest. The absolute risk is determined as the accumulation over the time interval [0, T] of the product between the event-free survival and the hazard of experiencing the event of interest, conditional to the baseline covariates. The event-free survival can be estimated from the cause-specific hazards using:

$$S(T|Z) = \exp\left(-\int_0^T \sum_{j=1}^J \lambda_j^{cs}(t|Z)\, dt\right) \quad (5)$$

where the hazard $\lambda_j^{cs}(t|Z)$ is termed proportional as it is the product of baseline hazard of its corresponding type of event $\lambda_{j0}^{cs}(t)$ and the corresponding function of the risk factors for the disease $\exp(\beta_j' Z)$.

Given the formula (4) and (5), it can be seen that the cause-specific CIF depends on the cause-specific hazard function for all the event types. In cause-specific hazards model, CIFs are estimated using Kaplan-Meier methods. While in Fine-Gray models, it models the effect of covariates on the CIF directly.

Fine-Gray model considers the subdistribution hazard function as [8]:

$$\lambda_j^{sh}(t|Z) = \lim_{\Delta t \to 0} \frac{P(T<t+\Delta t, J=j | Z\{T \geq t \text{ or } (T<t \text{ and } J \neq j)\})}{\Delta t} \quad (6)$$

*The absolute risk of the event of interest can be obtained by:*

$$F_1(T|Z) = 1 - \exp\left(\int_0^T \lambda_1^{sh}(t|Z) dt\right) \quad (7)$$

All above mentioned survival models reply on the validity of the proportional hazards assumption. A potential solution to the robustness problem is "stopped Cox regression", by using only the data that is directly associated to the survival probability *S(T|Z)* of interest [9]. One approach to address this is obtaining the survival probability provided that all observations are administratively censored at that target prediction horizon. This approach can be applied in the setting of competing risks, and also in the dynamic landmark models.

- ***Dynamic model***

The landmark approach in survival analysis involves selecting specific time points, known as "landmarks," at which the risk estimates for an event of interest are updated, using the information on the individuals who survive up to that given landmark time point.

Let *w* be the prediction window of interest. We aim to create a model to estimate risk at landmark time *s*, knowing an individual's covariates at *s*, namely *Z(s)*, conditioning on being alive at *s*. To create the landmark model, risk prediction times of interest are first partitioned into different landmarks $\{s_0, \ldots, s_L\}$. The sliding landmark datasets are created for each landmark *s*, using only the data of individuals still at risk at *s*, and applying artificial censoring at *s+w* to these individuals.

Separate Cox model can be fitted at each landmark *s* for which a prediction is required. However, this is impractical and difficult to communicate with clinical users [4]. Some form of smoothing and simplification is needed. This can be achieved by computing all the separate prediction models and smoothing them by constructing a stacked super dataset. In this case, we assume the baseline hazard depends on *s* and this can be modelled by $\lambda_0(t|Z(s), s) = \lambda_0(t) \exp(\gamma(s))$, *where* $\gamma(s) = \gamma_1 \left(\frac{s}{30}\right) + \gamma_2 \left(\frac{s}{30}\right)^2$. Note that $\frac{s}{30}$ is applied here to constrain the effect sizes of $\gamma_1$ and $\gamma_2$. The landmark Cox supermodel can be fitted by applying a Cox proportional hazards model to the stacked dataset of the different landmarks and the predicted survival function is:

$$S(s+w|Z(s), s) = \exp\left(-[\hat{\Lambda}(s+w|Z(s), s) - \hat{\Lambda}(s|Z(s), s)]\right)$$
$$= \exp\left(-\exp(\gamma(s) + \beta(s)Z(s))[\widehat{\Lambda_0}(s+w|Z(s), s) - \widehat{\Lambda_0}(s|Z(s), s)]\right) \quad (8)$$

where $\widehat{\Lambda_0}$ is the cumulative hazard and its numerical integration can be calculated using the *evalstep* function in *dynpred* package [10]. The predicted risk, that is, cumulative incidence, can be obtained using

$$F(s+w|Z(s),s) = 1 - S(s+w|Z(s),s) \tag{9}$$

Similarly, we can fit landmark cause-specific supermodel and consider the baseline hazards $\lambda_{j0}(t)$ from each of the cause-specific Cox models for event J (J=1,…, j) as $\lambda_{j0}^{cs}(t|Z(s),s) = \lambda_{j0}^{cs}(t)\exp\left(\gamma_j(s)\right)$ and $\gamma_j(s) = \gamma_{j1}\left(\frac{s}{30}\right) + \gamma_{j2}\left(\frac{s}{30}\right)^2$. Thus, the cause-specific hazards of supermodel for event J from a landmark time $s \in [s_0, s_L]$ for time $t$ ($s \leq t \leq s+w$) is:

$$\lambda_j^{cs}(t|Z(s),s) = \lambda_{j0}^{cs}(t)\exp\left(\gamma_j(s) + \beta_j(s)Z(s)\right) \tag{10}$$

Then *w*-year event free survival at any time point s in the window $[s_0, s_L]$ can be estimated with either the exponential approximation [11]:

$$S(s+w|Z(s),s) = \exp\left(-\int_s^{s+w}\sum_{j=1}^J \lambda_j^{cs}(t|Z(s),s)\,dt\right) = \exp\left(-\sum_{s\leq t_i \leq s+w}\sum_{j=1}^J \lambda_j^{cs}(t_i|Z(s),s)\right) \tag{11}$$

or the product integral estimator [11] when there are three competing events:

$$S(s+w|Z(s),s) = \prod_{s\leq t_i \leq s+w}(1 - d\Lambda_1^{cs}(t_i|Z(s),s) - d\Lambda_2^{cs}(t_i|Z(s),s) - d\Lambda_3^{cs}(t_i|Z(s),s)) \tag{12}$$

where $\Lambda_j^{cs}$ denotes the cumulative cause-specific hazard of event j. The cause-specific cumulative incidence for event J at any time point s in the window $[s_0, s_L]$ can be obtained with

$$F_j(s+w|Z(s),s) = \int_s^{s+w}\lambda_j^{cs}(t|Z(s),s)S(t|Z(s),s)\,dt = \sum_{s\leq t_i \leq s+w}\lambda_j^{cs}(t_i|Z(s),s)S(t_i|Z(s),s) \tag{13}$$

where $t_i$ are event times in the training dataset used to fitting the landmark cause-specific supermodel.

The concern in developing a landmark supermodel based on the Fine-Gray approach was in constructing the landmark dataset at each landmark to properly account for competing events that happened before the landmark in the setting of subdistribution hazard [12]. Liu et al. extended the landmark method to the Fine-Gray model and proposed landmark proportional subdistribution hazards (PSH) model [13] and the target dynamic prediction probabilities $F_j(s+w|Z(s),s)$ can be estimated as

$$F_j(s+w|Z(s),s) = 1 - \exp\left(-\left[\widehat{\Lambda_j^{sh}}(s+w|Z(s),s) - \widehat{\Lambda_j^{sh}}(s|Z(s),s)\right]\right)$$

$$= 1 - \exp\left(-\exp(\gamma_j(s) + \beta_j Z(s))\left[\widehat{\Lambda_{j0}^{sh}}(s+w|Z(s),s) - \widehat{\Lambda_{j0}^{sh}}(s|Z(s),s)\right]\right) \tag{14}$$

where $\widehat{\Lambda_j^{sh}}$ is the Breslow estimator for the cumulative subdistribution hazard of event J.

To implement the model in a Fine-Gray approach and obtain the estimated cumulative subdistribution hazard, each subset needs to be transformed into the counting process style before stacking all landmark subsets into a super dataset. Also, time-varying inverse probability censored weighting (IPCW) needs to be calculated for the subjects who experienced competing risks [14], which can be achieved with the function *finegray* from *survival* package [15].

As we mentioned in the baseline model, logistic regression can be used to link predictors to the binary event outcome. We used our landmark datasets which treats it as repeated observations at landmark times and related the probability of CLABSI occurring in an interval to a logistic function of the risk factors [16].

$$\log\left(\frac{P(Z(s),s)}{1-P(Z(s),s)}\right) = \beta_0 + \gamma_{log}(s) + \beta Z(s) \tag{15}$$

The parameter $\gamma_{log}$ denotes the effect of landmark time and here we assume a linear and quadratic trend on the time effect as $\gamma_{log}(s) = \gamma_1\left(\frac{s}{30}\right) + \gamma_2\left(\frac{s}{30}\right)^2$. The parameters $\beta_0$ and $\beta(s)$ are the intercept and variable estimates for the logistic landmark supermodel respectively.

Similar to the landmark logistic regression, we applied the multinomial logistic regression on the landmark dataset. The event indicator is not restricted to binary (0/1) cases. Here we used categorical variable to indicate the event (1=CLABSI; 2=Death, 3=Discharge/catheter removal; 0=administrative censored within 7 days) in each interval. Thus the probability of CLABSI can be obtained via

$$\log\left(\frac{P(Y=1|Z(s),s)}{P(Y=0|Z(s),s)}\right) = \beta_{10} + \gamma_{multi}(s) + \beta_1 Z(s) \tag{16}$$

The parameter $\gamma_{multi}$ denotes the effect of landmark time on CLABSI compared to base category and here we assume a linear and quadratic trend on the time effect as $\gamma_{multi}(s) = \gamma_{11}\left(\frac{s}{30}\right) + \gamma_{12}\left(\frac{s}{30}\right)^2$. The parameters $\beta_{10}$ and $\beta_1$ are like binary logistic regression expressing the relative log-odds of getting CLABSI compared to the base category.

Besides, we also implement regularized multi-task learning (MTL) for simultaneous learning of regression tasks on our stacked landmark datasets [17]. The algorithm follows the framework

$$\min_{W,C} \sum_{i=1}^{t} L(W_i, C_i | X_i, Y_i) + \lambda_1 \Omega(W) + \lambda_2 ||W||_F^2 \tag{17}$$

where X and Y are predictors matrices and responses of multiple tasks (landmarks) respectively. $L(W_i, C_i | X_i, Y_i)$ is the logistic loss function. W is the coefficient matrix. The algorithm incorporates not only the summation of logistic loss function across all landmarks, but also the cross-task regularization $\Omega(W)$ for knowledge transfer, and $||W||_F^2$ for improving the generalization. $\Omega(W)$ jointly modulates multi-tasks models $W_i$ across all landmarks according to the specific prior structure. In this case we used network-based relatedness, that is, $||WG||_F^2$ for $\Omega(W)$, which equals to an accumulation of differences between related tasks. The different of connected task can be 0 if the penalty is heavy enough.

Table S3 and Table S4 provided summaries of prediction algorithms for the landmark approach models in both baseline and dynamic models.

Table S3: summary of the algorithms for baseline models

| Model | Outcome | Prediction |
|---|---|---|
| binary logistic regression | $Y = \begin{cases} 1, & CLABSI \text{ occurred within 7 days}; \\ 0, & otherwise; \end{cases}$ | $\{1 + \exp[-\beta_0 - \beta'Z]\}^{-1}$ |
| multinomial logistic regression | $Y = \begin{cases} 1, & CLABSI \text{ occurred within 7 days}; \\ 2, & death \text{ occurred within 7 days}; \\ 3, & discharge/catheter\ removal \text{ within 7 days}; \\ 0, & otherwise; \end{cases}$ | $\frac{\exp(\beta_{0j}+\beta_j'Z)}{1+\sum_1^3 \exp(\beta_{0j}+\beta_j'Z)}$ with $j = 1,\dots,3$ |
| Cox proportional | [T, Y] where T is the failure time (when event occurs) and Y is the event indicator | $1 - \exp\left(-\int_0^7 \lambda_0(t)\exp(\beta'Z)\,dt\right)$ |
| hazards model | $Y = \begin{cases} 1, & CLABSI; \\ 0, & otherwise; \end{cases}$ | |
| cause-specific model | [T, Y] where T is the failure time (when event occurs) and Y is the event indicator<br>$Y = \begin{cases} 1, & CLABSI; \\ 2, & death; \\ 3, & discharge/catheter\ removal; \end{cases}$ | $\int_0^T \exp\left(-\int_0^7 \sum_{j=1}^J \lambda_j^{cs}(t\|Z)\,dt\right) \lambda_1^{cs}(t)dt$ |
| Fine-Gray model | [T, Y] where T is the failure time (when event occurs) and Y is the event indicator<br>$Y = \begin{cases} 1, & CLABSI; \\ 2, & death; \\ 3, & discharge/catheter\ removal; \end{cases}$ | $1 - exp\left(-\int_0^7 \lambda_1^{sh}(t\|Z)dt\right)$ |

Note: $Z' = (z_1, z_2, \ldots, z_k)$ is the set of $k$ risk factors, measured at the beginning (baseline) of follow-up period of length T. The parameter $\beta_0$ and $\beta$ is the intercept and coefficient estimates for the logistic model. $\beta_j$ is the row vector of regression coefficients of Z for the $j$th category of Y in multinomial logistic model. $\lambda_0(t)$ denotes the baseline hazard of Cox model, which refers to the probability that a person with all zero values for covariates will experience the event in the next instant if the person survives to $t$. Note that the coefficients vector $\beta$ for the risk factors $Z$ are not necessarily the same as the coefficients $\beta'$ in binary logistic model. $\lambda_j^{cs}(t)$ and $\lambda_j^{sh}(t)$ denotes the cause-specific and subdistribution hazard respectively for the corresponding event $j$.

Table S4: summary of the algorithms for dynamic models

| Model | Outcome | Prediction |
|---|---|---|
| binary logistic landmark supermodel | $Y = \begin{cases} 1, & CLABSI\ occurred\ within\ s+w\ days; \\ 0, & otherwise; \end{cases}$ | $\log\left(\frac{P(Z(s),s)}{1-P(Z(s),s)}\right) = \beta_0 + \gamma_{log}(s) + \beta Z(s)$ |
| regularized multi-task learning (RMTL) | $Y = \begin{cases} 1, & CLABSI\ occurred\ within\ s+w\ days; \\ -1, & otherwise; \end{cases}$ | $\log\left(\frac{P(Z(s))}{1-P(Z(s))}\right) = w_s Z^*(s)$ where $w_s$ are the coefficients at each landmark s and $Z^*(s)$ represents the standardized covariates at s |
| multinomial logistic landmark | $Y = \begin{cases} 1, & CLABSI\ occurred\ within\ s+w\ days; \\ 2, & death\ occurred\ within\ s+w\ days; \\ 3, & discharge/catheter\ removal\ within\ s+w\ days; \\ 0, & otherwise; \end{cases}$ | $\log\left(\frac{P(Y=1\|Z(s),s)}{P(Y=0\|Z(s),s)}\right) = \beta_{10} + \gamma_{multi}(s) + \beta_1 Z(s)$ |
| Cox landmark supermodel | [T*, Y] where T is the administratively censored failure time T* = min(T, C) where C is the administrative censoring time ($s+w$) and Y is the event indicator<br>$Y = \begin{cases} 1, & CLABSI\ within\ s+w\ days; \\ 0, & otherwise; \end{cases}$ | $1 - \exp(-[\widehat{\Lambda}(s+w\|Z(s),s) - \widehat{\Lambda}(s\|Z(s),s)])$ |
| cause-specific landmark supermodel | [T*, Y] where T is administratively censored failure time T* = min(T, C) where C is the administrative censoring time ($s+w$) and Y is the event indicator<br>$Y = \begin{cases} 1, & CLABSI\ within\ s+w\ days; \\ 2, & death\ within\ s+w\ days; \\ 3, & discharge/catheter\ removal\ within\ s+w\ days; \\ 0, & otherwise \end{cases}$ | $\sum_{s \le t_i \le s+w} \lambda_1^{cs}(t_i\|Z(s),s) S(t_i\|Z(s),s)$ |
| Fine-Gray landmark supermodel | [T*, Y] where T is administratively censored failure time T* = min(T, C) where C is the administrative censoring time ($s+w$) Y is the event indicator<br>$Y = \begin{cases} 1, & CLABSI\ within\ s+w\ days; \\ 2, & death\ within\ s+w\ days; \\ 3, & discharge/catheter\ removal\ within\ s+w\ days; \\ 0, & otherwise \end{cases}$ | $1 - \exp\left(-\left[\widehat{\Lambda_1^{sh}}(s+w\|Z(s),s) - \widehat{\Lambda_1^{sh}}(s\|Z(s),s)\right]\right)$ |

Note: In logistic landmark supermodels, the parameter $\gamma$ denotes the effect of landmark time and here we assume a linear and quadratic trend on the time effect as $\gamma_{log}(s) = \gamma_1 \left(\frac{s}{30}\right) + \gamma_2 \left(\frac{s}{30}\right)^2$. The parameters $\beta_0$ and $\beta$ are the intercept and variable estimates for the logistic landmark supermodel respectively. Similarly, $\gamma_{multi}(s)$ denotes the effect of landmark time on CLABSI compared to base category (censored cases) and here we assume a linear and quadratic trend on the time effect as $\gamma_{multi}(s) = \gamma_{11} \left(\frac{s}{30}\right) + \gamma_{12} \left(\frac{s}{30}\right)^2$. The parameters $\beta_{10}$ and $\beta_1$ are like binary logistic regression expressing the relative log-odds of getting CLABSI compared to the base category. In survival framework, we assume the baseline hazard depends on $s$ and can be modelled by $\lambda_0(t|Z(s),s) = \lambda_0(t) \exp(\gamma(s))$, $where \, \gamma(s) = \gamma_1 \left(\frac{s}{30}\right) + \gamma_2 \left(\frac{s}{30}\right)^2$. Note that $\frac{s}{30}$ is applied here to constrain the effect sizes of $\gamma_1$ and $\gamma_2$. $\widehat{\Lambda}$ is the cumulative hazard when there are no competing risks, and $\widehat{\Lambda_j^{sh}}$ is the Breslow estimator for the cumulative subdistribution hazard of event J in Fine-Gray framework. In cause-specific landmark supermodel, the cause-specific hazards for event J from a landmark time $s \in [s_0, s_L]$ for time $t$ ($s \leq t \leq s+w$) is $\lambda_j^{cs}(t|Z(s),s)$ and can be obtained via $\lambda_{j0}^{cs}(t)\exp(\gamma_j(s) + \beta_j(s)Z(s))$ where $\lambda_{j0}^{cs}(t)$ denotes the baseline hazards from each of the cause-specific models for event J (J=1,2,3) and $\gamma_j(s)$ also assumes a linear and quadratic effect on $\left(\frac{s}{30}\right)$.

# Supplementary file 4: missing data

Table S5: missing data

| Column | Description | Variable | Missing at LM0 | Missing at all LMs |
|---|---|---|---|---|
| Central venous catheter (CVC) | Was there a catheter of type CVC connected since previous LM? | time-varying | 0.0% | 0.0% |
| Port a cath | Was there a catheter of type Port_a_cath connected since previous LM? | time-varying | 0.0% | 0.0% |
| Tunneled central venous catheter | Was there a catheter of type tunneled CVC connected since previous LM | time-varying | 0.0% | 0.0% |
| Peripherally inserted central catheter (PICC) | Was there a catheter of type PICC connected since previous LM? | time-varying | 0.0% | 0.0% |
| Subclavian | Was a catheter connected in the Subclavian location since previous LM? | time-varying | 0.0% | 0.0% |
| Jugular | Was a catheter connected in the Jugular location since previous LM? | time-varying | 0.0% | 0.0% |
| Total parental nutrition (TPN) | Has TPN been ordered for the patient in the previous 7 days from LM time | time-varying | 0.0% | 0.0% |
| Antineoplastic agents | Have any drugs in ATC group (level 2) L01 (ANTINEOPLASTIC AGENTS) been ordered for the patient in the previous 7 days from LM time | time-varying | 0.0% | 0.0% |
| Antibacterials for systematic use | Have any drugs in ATC group (level 2) J01 (ANTIBACTERIALS FOR SYSTEMIC USE) been ordered for the patient in the previous 7 days from LM time | time-varying | 0.0% | 0.0% |
| CLABSI history | Did the patient experience a CLABSI event in the past 3 months since LM time? | time-varying | 0.0% | 0.0% |
| Tumor | Has a tumour pathology been registered before current LM time? | time-varying | 0.0% | 0.0% |
| Lymphoma | Has lymphoma been registered as a comorbidity before current LM time? | time-varying | 0.0% | 0.0% |
| Transplant | Has a transplant pathology been registered before current LM time? | time-varying | 0.0% | 0.0% |
| ICU unit | Is the patient now (at the exact second of the current LM) in ICU? | time-varying | 0.0% | 0.0% |
| Mechanical ventilation (MV) | Is the patient on MV since previous landmark? A patient is considered on MV if at least one value of PEEP or FiO2 are recorded between 2 landmarks. Only valid for ICU patients. | time-varying | 0.0% | 0.0% |
| Temperature | Maximum value of temperature since previous landmark. For baseline (LM 0) the last value from the previous 24 hours is used. Only temperatures in the range (30 °C, 45 °C) are kept, the others are filtered. Maximum value is used to correct for very low temperatures measured by devices in ICU, when the temperature falls closer to the room temperature. | time-varying | 26.7% | 10.9% |
| Systolic blood pressure | Last value of systolic blood pressure since previous landmark. For baseline (LM 0) the last value from the previous 24 hours is used. | time-varying | 28.0% | 15.3% |
| White blood cells (WBC) count | WBC count, last value since previous LM. Unit: 10**9/L | time-varying | 41.7% | 39.4% |
| C-reactive protein (CRP) | CRP, last value since previous LM. Unit: mg/L | time-varying | 46.7% | 40.5% |
| Positive culture, of any other type than blood | Has there been a positive culture, of any other type than blood, in the last 17 days (time window used for secondary BSIs). | time-varying | 0.0% | 0.0% |
| Patients admitted from home | Was the patient admitted from home? | baseline | 1.6% | not applicable |

# Supplementary file 5: sample size calculation

We used the pmsampsize package in R to estimate the minimum sample size required for developing a static logistic regression model and a static Cox proportional hazards [18]. According to a systematic review of CLABSI prediction models [19], the mean of optimism-corrected AUC is 0.75 in the studies with similar EHR setting. Assuming we can obtain a c-statistic of 0.75 with 21 parameters and a prevalence of 3.14%, a minimum of 7,027 and 6,974 unique catheter episodes are required to develop logistic and Cox models, respectively. When the prevalence is 1.31%, at least 16,621 (logistic) and 15,912 (Cox) catheter episodes are required. Hence, our sample size of 30,862 is sufficiently large for static models if applying a 2:1 train test split. Given the utilization of the landmark approach for dynamic models, where information from adjacent landmarks is leveraged, we conclude that the same sample size rationale is applicable for dynamic models as well.

# Supplementary file 6: baseline models performances

Figure S1: AUC plots of all baseline models for 100 train-test splits

(a) Cox-ac

(b) Cox

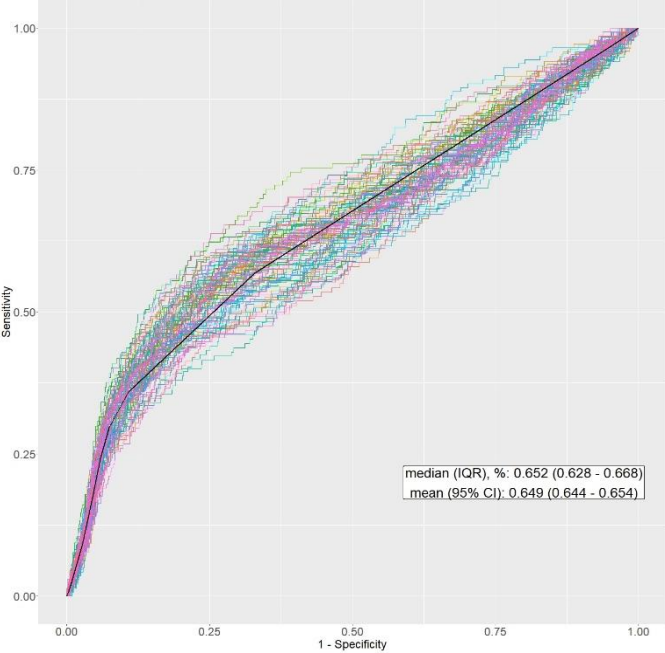
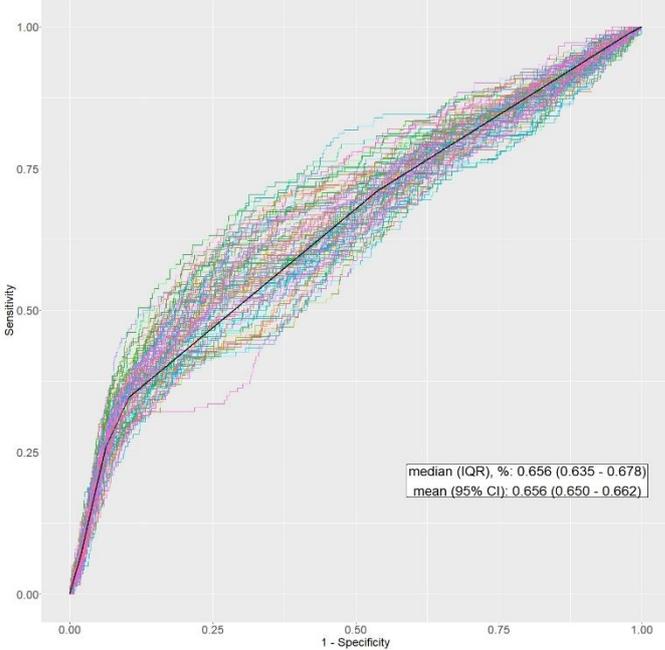

(c) CS-ac

(d) CS

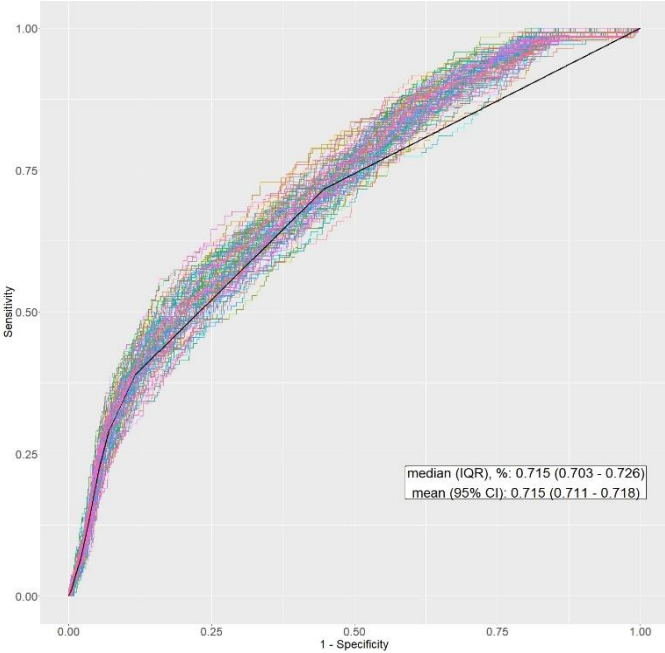
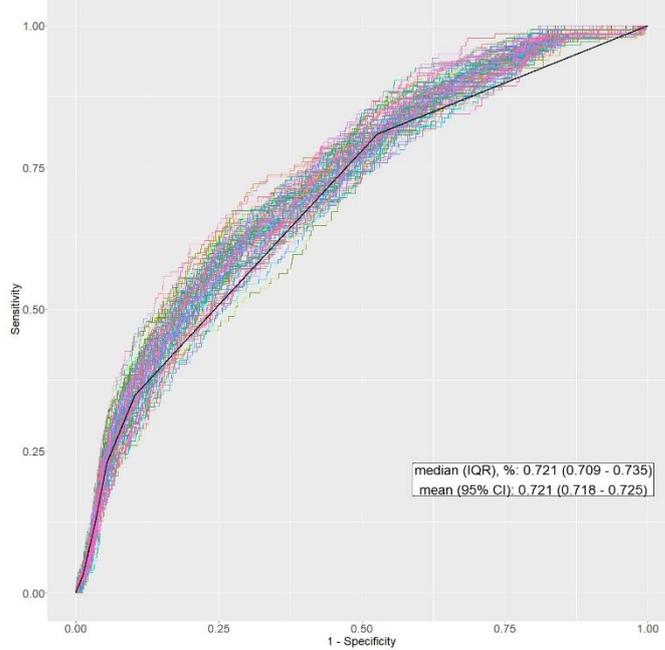

(e) FG-ac

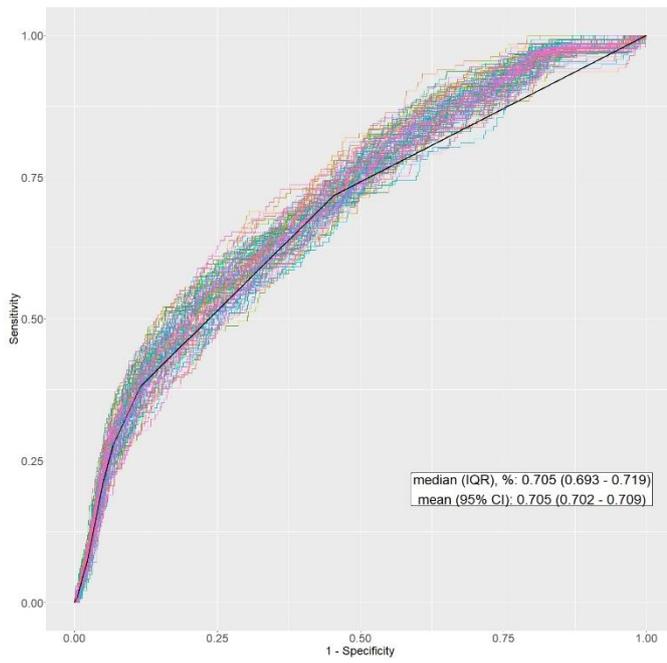

(f) FG

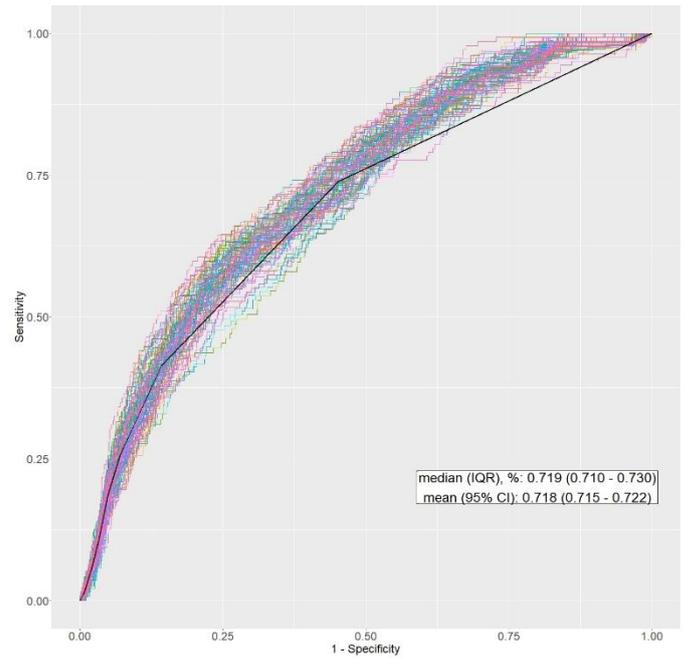

(g) LG

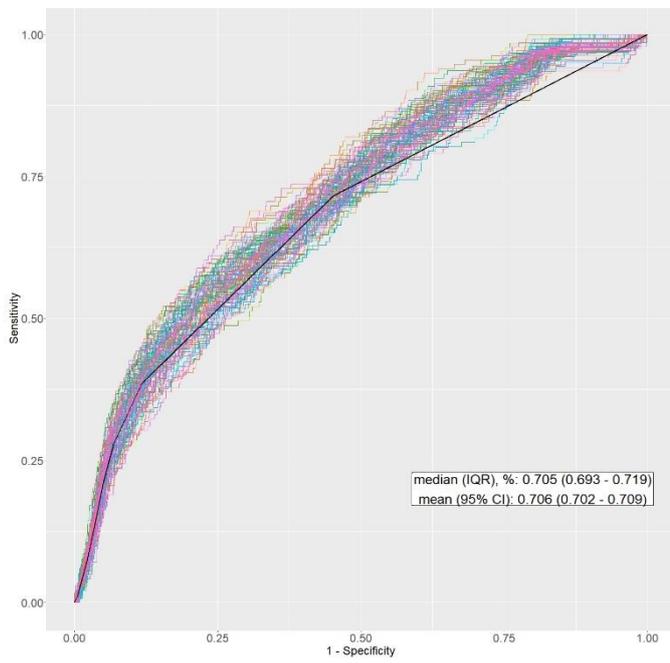

(h) MLR

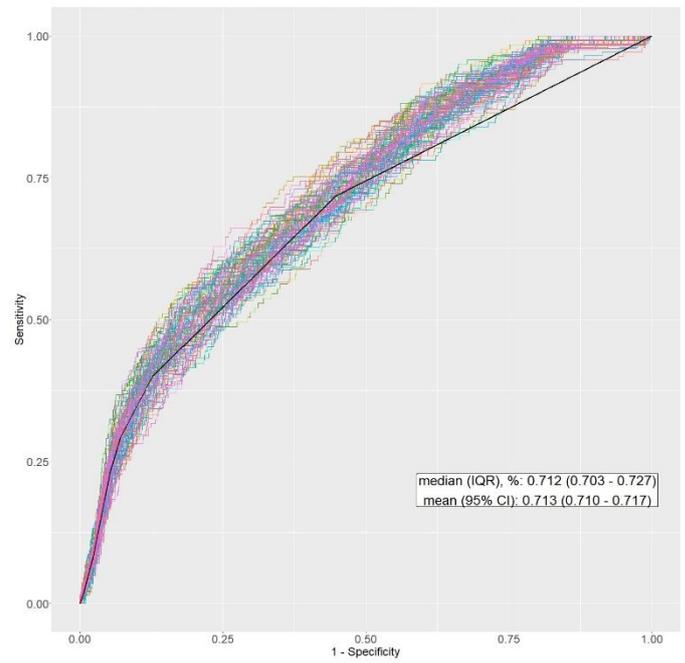

Figure S2: Calibration plots of all baseline models for 100 train-test splits

(a) Cox-ac

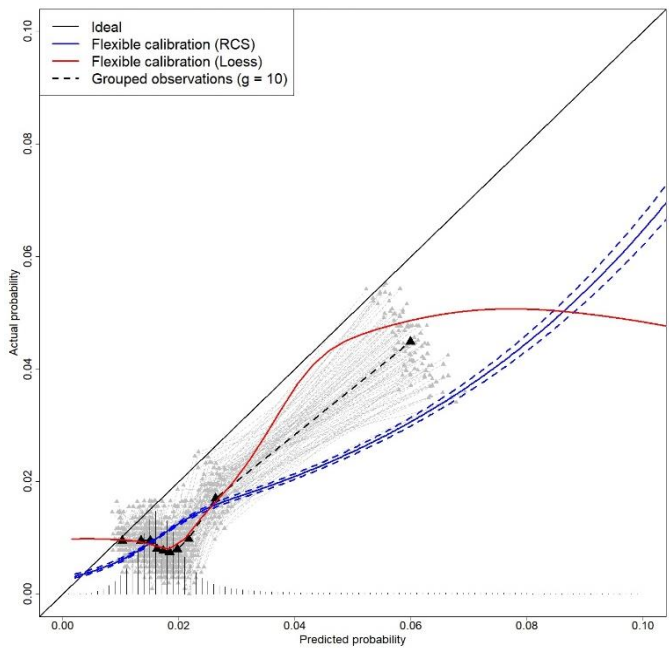

(b) Cox

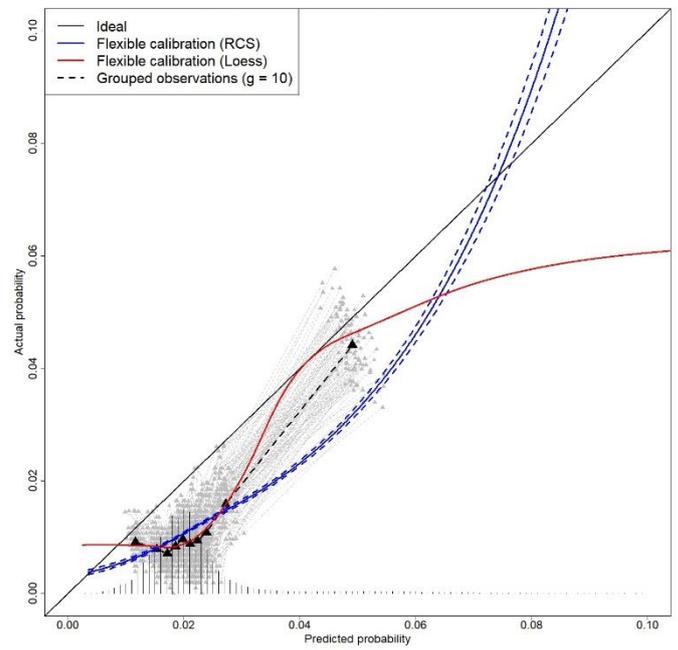

(c) CS-ac

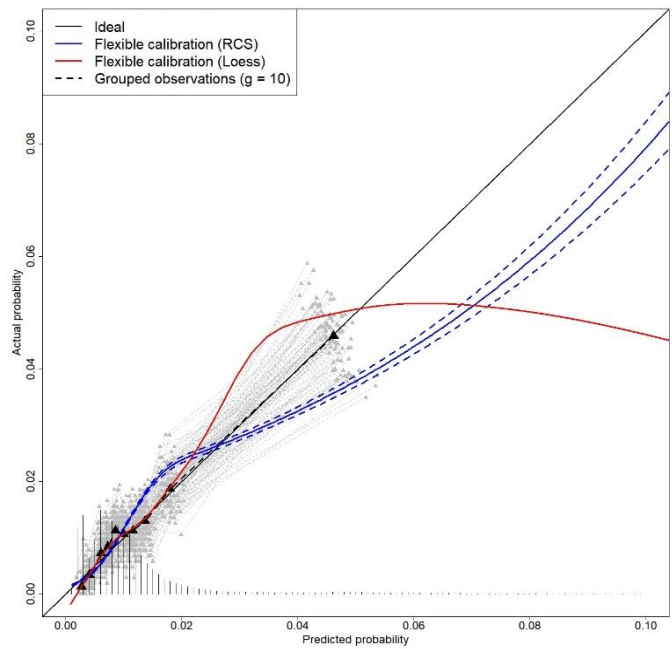

(d) CS

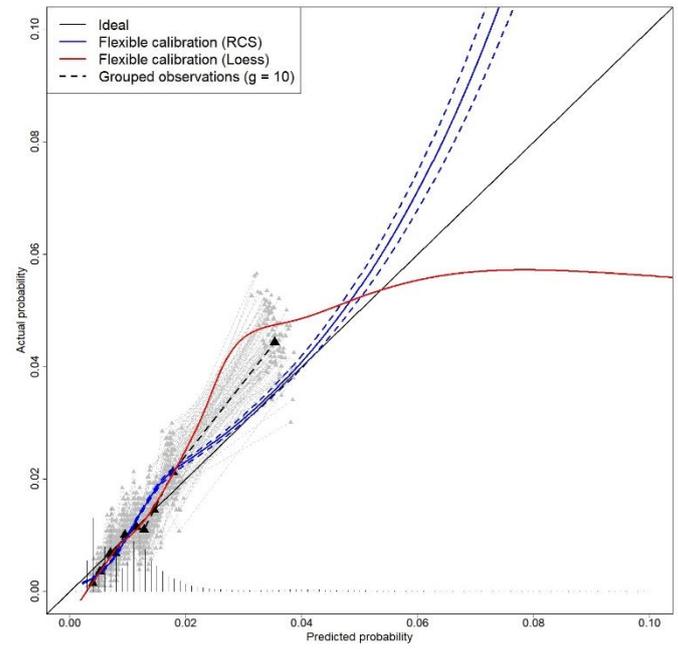

(e) FG-ac
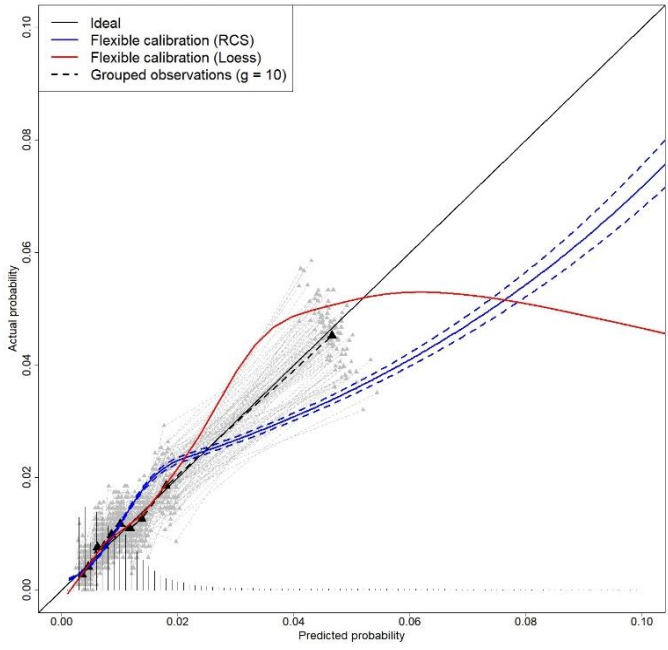

(f) FG
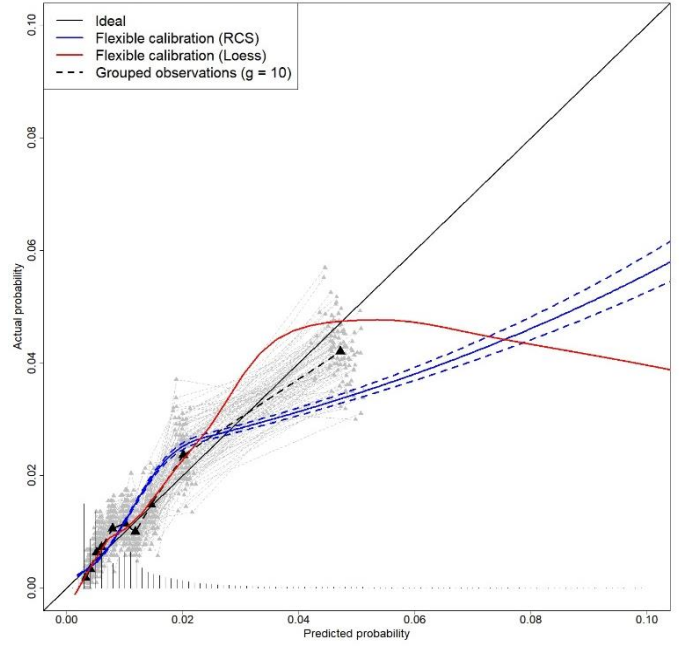

(g) LG
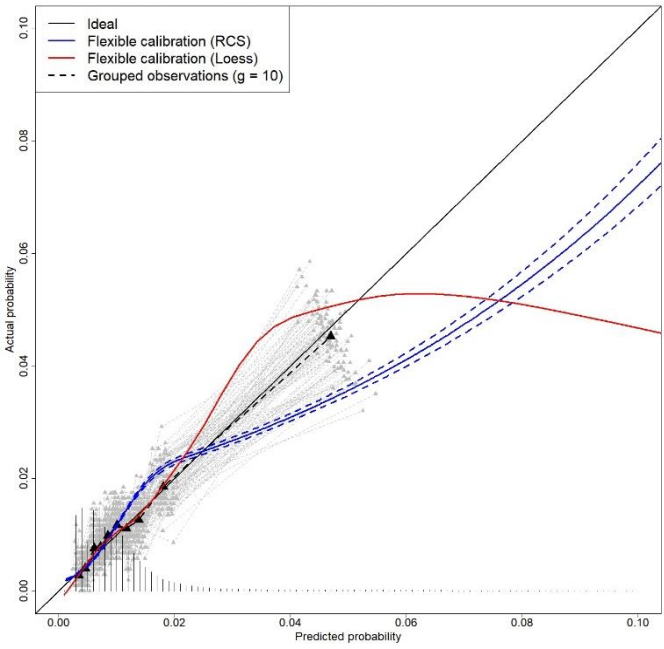

(h) MLR
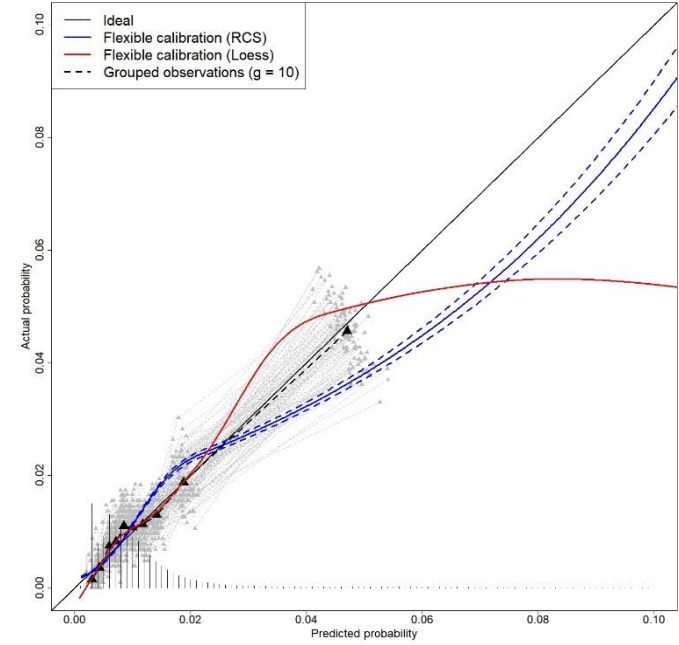

# Supplementary file 7: separate fine-gray landmark model

We fitted separate fine-gray models to our landmark super datasets. It shall be noticed that convergence started failure since landmark day 14 for separate Fine-Gray landmark model. The number of convergence failures is shown in Table S6.

Table S6: number of convergence failure per landmark for separate Fine-Gray landmark model

| LM | Nr of convergence failure (fg_model$crrFit$converged==FALSE) |
|----|---|
| 30 | 99 |
| 29 | 99 |
| 28 | 100 |
| 27 | 100 |
| 26 | 100 |
| 25 | 98 |
| 24 | 100 |
| 23 | 100 |
| 22 | 100 |
| 21 | 100 |
| 20 | 17 |
| 19 | 16 |
| 18 | 12 |
| 17 | 12 |
| 16 | 12 |
| 15 | 7 |
| 14 | 1 |

# Supplementary file 8: coefficient estimates

Figure S3: Coefficient estimates of the baseline models

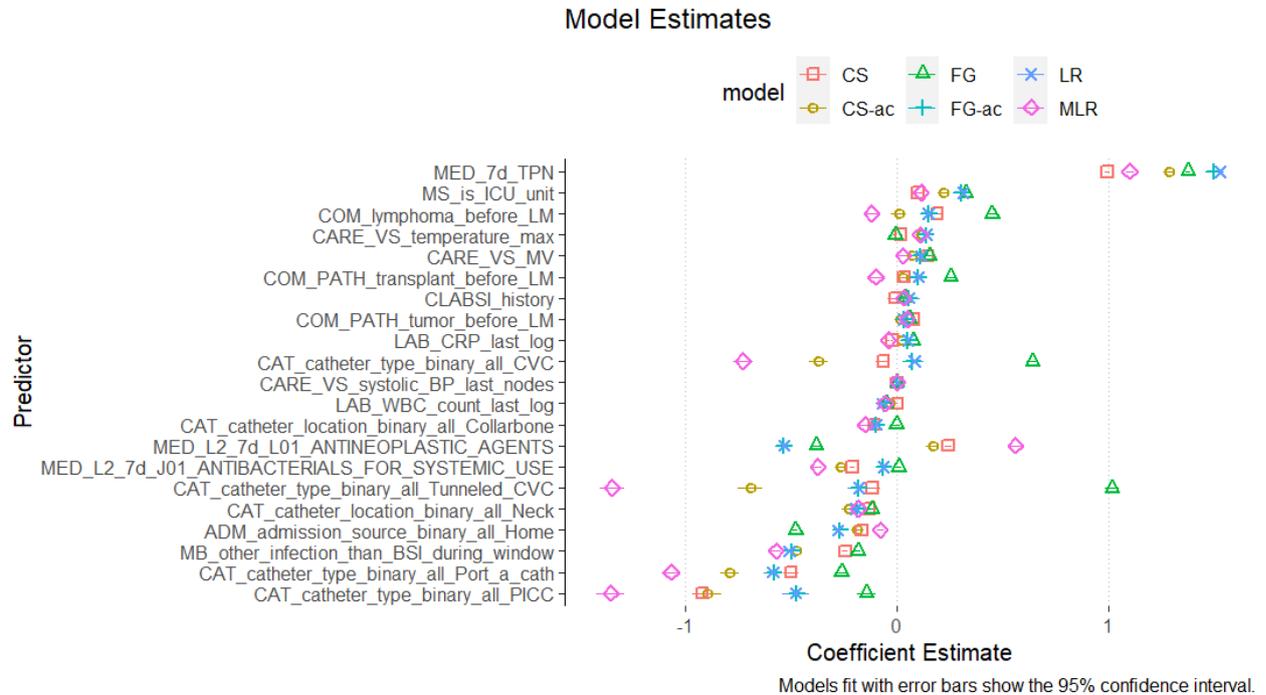

Note: the coefficient estimates of cox models (administrative censoring at day 7 and without administrative censoring) are not shown as they are the same as cause-specific models.

Table S7: coefficient estimates of baseline models

|  | CS-ac | CS | FG-ac | FG | LR | MLR |
|---|---|---|---|---|---|---|
| Patients admitted from home | -0.187 (-0.204, -0.17) | -0.166 (-0.178, -0.155) | -0.27 (-0.287, -0.253) | -0.479 (-0.491, -0.468) | -0.279 (-0.296, -0.261) | -0.08 (-0.097, -0.062) |
| Mechanical ventilation | 0.071 (0.044, 0.098) | 0.142 (0.124, 0.16) | 0.108 (0.081, 0.135) | 0.154 (0.137, 0.172) | 0.112 (0.084, 0.139) | 0.029 (0.002, 0.056) |
| Systolic blood pressure | 0.001 (0, 0.001) | 0 (0, 0) | 0.001 (0, 0.001) | -0.001 (-0.001, -0.001) | 0.001 (0, 0.001) | 0.001 (0, 0.001) |
| Temperature | 0.114 (0.104, 0.124) | 0.018 (0.012, 0.025) | 0.135 (0.125, 0.145) | -0.008 (-0.015, -0.001) | 0.137 (0.127, 0.148) | 0.111 (0.1, 0.121) |
| Catheter location: Subclavian | -0.107 (-0.138, -0.077) | -0.109 (-0.13, -0.087) | -0.099 (-0.128, -0.07) | -0.003 (-0.024, 0.018) | -0.097 (-0.127, -0.067) | -0.151 (-0.183, -0.12) |
| Catheter location: Jugular | -0.23 (-0.259, -0.201) | -0.131 (-0.153, -0.109) | -0.191 (-0.218, -0.164) | -0.118 (-0.138, -0.098) | -0.195 (-0.223, -0.168) | -0.182 (-0.211, -0.154) |
| Catheter type: CVC | -0.371 (-0.417, -0.325) | -0.065 (-0.099, -0.032) | 0.073 (0.03, 0.116) | 0.64 (0.612, 0.669) | 0.081 (0.037, 0.125) | -0.728 (-0.775, -0.681) |
| Catheter type: PICC | -0.896 (-0.96, -0.831) | -0.922 (-0.966, -0.878) | -0.476 (-0.537, -0.415) | -0.145 (-0.187, -0.102) | -0.479 (-0.541, -0.416) | -1.353 (-1.42, -1.286) |
| Catheter type: Port_a_cath | -0.791 (-0.837, -0.745) | -0.502 (-0.539, -0.466) | -0.582 (-0.625, -0.54) | -0.263 (-0.294, -0.231) | -0.586 (-0.629, -0.544) | -1.068 (-1.114, -1.022) |
| Catheter type: Tunneled CVC | -0.694 (-0.752, -0.637) | -0.116 (-0.153, -0.079) | -0.183 (-0.235, -0.131) | 1.016 (0.985, 1.047) | -0.18 (-0.233, -0.127) | -1.347 (-1.404, -1.29) |
| CLABSI history | 0.051 (0, 0.103) | -0.01 (-0.045, 0.025) | 0.055 (0.004, 0.107) | 0.037 (0.004, 0.07) | 0.056 (0.004, 0.109) | 0.031 (-0.021, 0.083) |
| Lymphoma | 0.009 (-0.026, 0.044) | 0.189 (0.168, 0.21) | 0.149 (0.114, 0.183) | 0.448 (0.427, 0.47) | 0.148 (0.113, 0.183) | -0.121 (-0.157, -0.085) |
| Transplant | 0.029 (-0.002, 0.06) | 0.033 (0.014, 0.051) | 0.099 (0.067, 0.13) | 0.254 (0.235, 0.273) | 0.101 (0.069, 0.133) | -0.101 (-0.133, -0.069) |
| Tumor | 0.017 (-0.004, 0.038) | 0.078 (0.064, 0.091) | 0.032 (0.01, 0.054) | 0.06 (0.046, 0.074) | 0.029 (0.007, 0.051) | 0.047 (0.025, 0.069) |
| CRP (unit: mg/L) | 0.021 (0.016, 0.027) | -0.023 (-0.027, -0.02) | 0.051 (0.046, 0.057) | 0.077 (0.073, 0.081) | 0.052 (0.047, 0.058) | -0.041 (-0.047, -0.035) |
| WBC count (unit: $10^9$/L) | -0.037 (-0.051, -0.022) | -0.002 (-0.01, 0.006) | -0.063 (-0.078, -0.048) | -0.053 (-0.062, -0.044) | -0.066 (-0.082, -0.051) | -0.055 (-0.071, -0.04) |
| Positive culture | -0.474 (-0.497, -0.452) | -0.245 (-0.258, -0.231) | -0.499 (-0.521, -0.476) | -0.186 (-0.199, -0.172) | -0.51 (-0.533, -0.487) | -0.572 (-0.595, -0.55) |
| TPN | 1.29 (1.272, 1.308) | 0.995 (0.984, 1.006) | 1.495 (1.476, 1.513) | 1.376 (1.365, 1.387) | 1.527 (1.508, 1.546) | 1.1 (1.082, 1.119) |
| Antibacterials for systematic use | -0.265 (-0.287, -0.244) | -0.211 (-0.223, -0.2) | -0.065 (-0.087, -0.042) | 0.008 (-0.005, 0.021) | -0.064 (-0.087, -0.041) | -0.375 (-0.397, -0.352) |
| Antineoplastic agents | 0.17 (0.14, 0.199) | 0.243 (0.226, 0.261) | -0.537 (-0.57, -0.505) | -0.383 (-0.403, -0.363) | -0.534 (-0.567, -0.502) | 0.559 (0.528, 0.591) |
| ICU_unit | 0.221 (0.2, 0.242) | 0.098 (0.084, 0.111) | 0.301 (0.28, 0.322) | 0.326 (0.312, 0.339) | 0.31 (0.289, 0.332) | 0.113 (0.092, 0.135) |

Figure S4: Coefficient estimates of the dynamic models

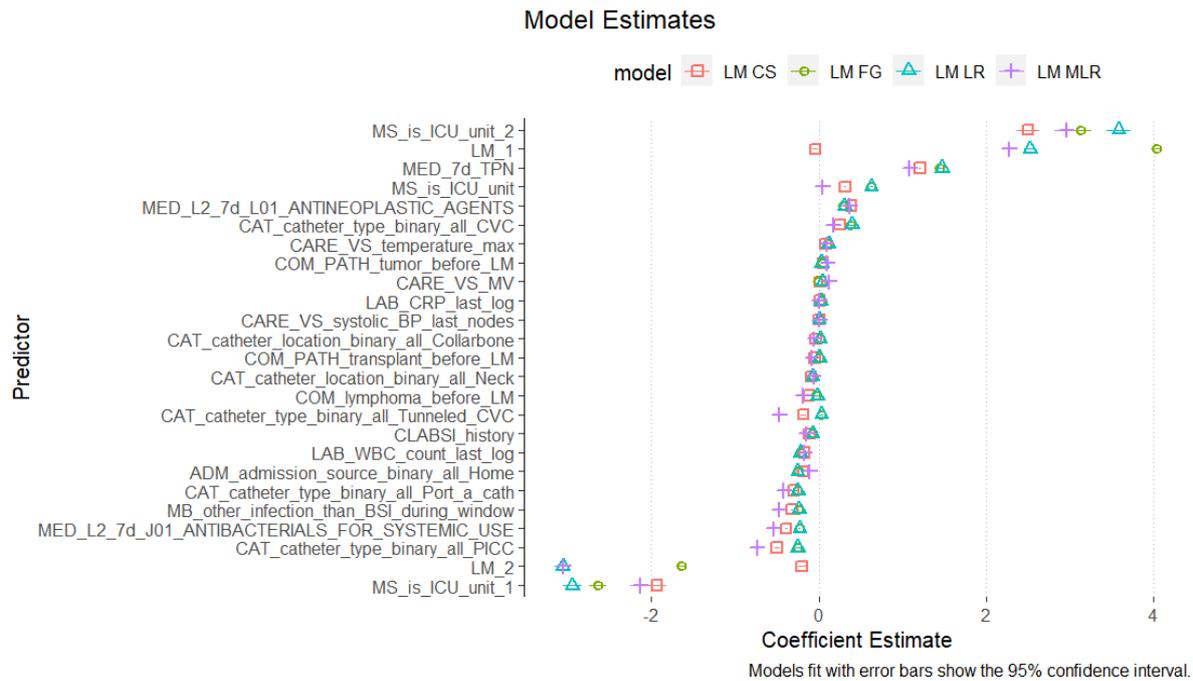

Note: the coefficient estimates of separate logistic landmark model and RMTL model are not shown as they have different coefficients across all landmarks; the coefficient estimates of cox landmark supermodels is not shown as they are the same as cause-specific landmark supermodel.

Table S8: coefficient estimates of dynamic models

| | LM CS | LM FG | LM LR | LM MLR |
|---|---|---|---|---|
| Patients admitted from home | -0.183 (-0.196, -0.17) | -0.245 (-0.258, -0.232) | -0.256 (-0.269, -0.243) | -0.108 (-0.121, -0.094) |
| Mechanical ventilation | 0.019 (0.007, 0.03) | -0.003 (-0.014, 0.008) | 0.041 (0.03, 0.053) | 0.119 (0.108, 0.131) |
| Systolic blood pressure | 0.002 (0.002, 0.002) | 0.003 (0.002, 0.003) | 0.003 (0.003, 0.003) | 0.002 (0.002, 0.003) |
| Temperature | 0.085 (0.08, 0.089) | 0.112 (0.107, 0.117) | 0.118 (0.113, 0.123) | 0.091 (0.086, 0.096) |
| Catheter location: Subclavian | -0.034 (-0.048, -0.02) | 0.012 (-0.001, 0.026) | 0.015 (0.001, 0.028) | -0.055 (-0.069, -0.04) |
| Catheter location: Jugular | -0.094 (-0.109, -0.078) | -0.074 (-0.089, -0.06) | -0.074 (-0.089, -0.059) | -0.064 (-0.08, -0.048) |
| Catheter type: CVC | 0.252 (0.227, 0.277) | 0.379 (0.355, 0.403) | 0.398 (0.373, 0.422) | 0.167 (0.142, 0.192) |
| Catheter type: PICC | -0.499 (-0.533, -0.464) | -0.259 (-0.293, -0.225) | -0.252 (-0.286, -0.218) | -0.741 (-0.775, -0.706) |
| Catheter type: Port_a_cath | -0.302 (-0.328, -0.276) | -0.257 (-0.281, -0.232) | -0.253 (-0.278, -0.227) | -0.418 (-0.443, -0.393) |
| Catheter type: Tunneled CVC | -0.186 (-0.214, -0.157) | 0.028 (0.001, 0.055) | 0.031 (0.003, 0.059) | -0.477 (-0.506, -0.449) |
| CLABSI history | -0.116 (-0.15, -0.081) | -0.079 (-0.113, -0.045) | -0.078 (-0.113, -0.043) | -0.15 (-0.186, -0.114) |
| Lymphoma | -0.12 (-0.143, -0.097) | -0.035 (-0.058, -0.011) | -0.018 (-0.042, 0.006) | -0.189 (-0.214, -0.165) |
| Transplant pathology | -0.051 (-0.071, -0.03) | -0.001 (-0.022, 0.019) | 0.007 (-0.015, 0.028) | -0.084 (-0.105, -0.064) |
| Tumor pathology | 0.047 (0.034, 0.06) | 0.037 (0.024, 0.051) | 0.035 (0.021, 0.048) | 0.098 (0.085, 0.112) |
| CRP (unit: mg/L) | 0.01 (0.007, 0.014) | 0.024 (0.02, 0.027) | 0.028 (0.024, 0.032) | 0.006 (0.003, 0.01) |
| WBC count (unit: $10^9$/L) | -0.171 (-0.177, -0.166) | -0.213 (-0.219, -0.208) | -0.22 (-0.226, -0.214) | -0.171 (-0.177, -0.166) |
| landmark/30 | -0.043 (-0.077, -0.01) | 4.04 (3.999, 4.082) | 2.521 (2.449, 2.593) | 2.272 (2.199, 2.344) |
| (landmark/30)^2 | -0.201 (-0.246, -0.157) | -1.639 (-1.692, -1.585) | -3.06 (-3.154, -2.966) | -3.049 (-3.144, -2.954) |
| Positive culture | -0.322 (-0.332, -0.312) | -0.236 (-0.246, -0.227) | -0.24 (-0.25, -0.23) | -0.481 (-0.492, -0.471) |
| TPN | 1.208 (1.197, 1.219) | 1.44 (1.429, 1.451) | 1.473 (1.461, 1.484) | 1.085 (1.073, 1.096) |
| Antibacterials for systematic use | -0.391 (-0.404, -0.378) | -0.232 (-0.245, -0.219) | -0.229 (-0.242, -0.216) | -0.544 (-0.558, -0.53) |
| Antineoplastic agents | 0.388 (0.373, 0.404) | 0.283 (0.268, 0.299) | 0.304 (0.288, 0.32) | 0.364 (0.348, 0.38) |
| ICU_unit | 0.314 (0.296, 0.332) | 0.632 (0.614, 0.651) | 0.625 (0.607, 0.644) | 0.045 (0.027, 0.063) |
| ICU_unit * landmark/30 | -1.93 (-2.037, -1.823) | -2.642 (-2.749, -2.534) | -2.947 (-3.061, -2.834) | -2.134 (-2.248, -2.02) |
| ICU_unit * (landmark/30)^2 | 2.5 (2.369, 2.63) | 3.131 (3, 3.262) | 3.582 (3.439, 3.725) | 2.956 (2.811, 3.1) |

# Supplementary file 9: codes

The codes used for model building, prediction and evaluation can be accessed via: https://gitlab.kuleuven.be/u0112758/clabsi_compare_models